\documentclass[twocolumn,preprintnumbers,amsmath,amssymb,superscriptaddress]{revtex4-2}
\UseRawInputEncoding

\usepackage{latexsym,amssymb,amsthm,amsmath,epsfig,braket}
\usepackage[left=2.00cm, right=2.00cm, top=2.00cm, bottom=2.00cm]{geometry}
\usepackage{xcolor}
\usepackage{svg}
\usepackage[colorlinks, citecolor={blue!80!black}, urlcolor={blue!80!black}, linkcolor={red!50!black}]{hyperref}
\usepackage{float}
\usepackage{hyperref}
\hypersetup{
    citecolor=red,
    colorlinks=true,
    linkcolor=blue,
    filecolor=blue,      
    urlcolor=blue,
}

\begin{document}

\title{Topological transport in monolayer jacutingaite}

\author{Muzamil Shah}
\email{muzamill@zjnu.edu.cn}
\affiliation{Department of Physics, Zhejiang Normal University, Jinhua, Zhejiang 321004, China}
\affiliation{Zhejiang Institute of Photoelectronics \& Zhejiang Institute for Advanced Light Source, Zhejiang Normal University, Jinhua, Zhejiang 321004, China}
\affiliation{Department of Physics, Quaid-I-Azam University Islamabad, 45320, Pakistan}

\author{Muhammad Sabieh Anwar}
\email{sabieh@lums.edu.pk}
\affiliation{Department of Physics, Syed Babar Ali School of Science and Engineering, Lahore University of Management Sciences (LUMS), Opposite Sector U, D.H.A., Lahore 54792, Pakistan \& Temel Bilimler Ara\c{s}t{\i}rma Enstit\"{u}s\"{u} (TBAE), T\"{U}B\.{I}TAK, Gebze, Turkey}

\date{\today}

\begin{abstract}
	Monolayer-jacutingaite ($\mathrm{Pt_{2}HgSe_{3}}$) has been predicted to be the first large-gap Kane-Mele quantum spin Hall insulator. Materials in the jacutingaite family undergo topological phase transitions (TPTs), \emph{i.e.}, from a topologically non-trivial to a semimetallic phase and further to the normal insulating phase when exposed to electric fields and off-resonance, high-frequency and high-intensity laser irradiation. In this article, we investigate the rich tapestry of topological phases in this unique material in the presence of an appropriate choice of off-resonance circularly polarized laser fields and staggered sublattice potentials. The interplay of these stimuli with large spin-orbit coupling, due to the buckled structure of jacutingaite materials,  results in the emergence of quantum spin Hall insulator, valley-spin-polarized metal, spin-polarized metal, photo-induced quantum Hall insulator, anomalous quantum Hall insulator and band insulator phases.  By analyzing the band structures, we compute Berry curvatures in different topological regimes for the $K$ and $K'$ valleys.  Furthermore, by using the Kubo formula, we calculate the spin-valley resolved longitudinal and  Hall conductivities as a function of photon energies showing that the conductivities exhibit a strong topological state dependence.  The photon energy of the intraband and interband optical transitions can be tuned by varying the electric and optical fields. Finally, we demonstrate that by modulating the chemical potential, some of the allowed optical transitions become Pauli blocked due to the optical selection rules.
	
\end{abstract}

\maketitle

\section{ Introduction}
Novel topological phases have attracted immense interest \cite{hasan2010colloquium,qi2011topological} and are hailed as a paradigm shift in the realm of condensed matter physics. One of the hallmarks of topological matter is the appearance of edge states and spin-momentum locking in two-dimensional (2D) Dirac systems. These phases are robust to backscattering at the interface and carry dissipationless currents \cite{bernevig2013topological}. In this context, quantum spin-Hall insulators (QSHI) also known as 2D topological insulators are a novel electronic state originating from spin-orbit coupling (SOC) and possess topological edge states that are also protected by time-reversal symmetry ($\mathcal{T}$) \cite{PhysRevLett.95.226801, PhysRevLett.95.146802}. Helical gapless edge states are characterized by counter-propagating spin-momentum locking with zero net electronic conductance due to the equal and opposite contributions from the two spin channels \cite{bernevig2006quantum}. In 2D hexagonal crystalline structures, the spin-orbit coupling is the 
key characteristic to the manifestation of the QSHI state as predicted by Kane and Mele \cite{PhysRevLett.95.226801}. Materials from the graphene family, overall referred to as 2D buckled X-enes (silicene, germanene, plumbene, and tinene) are notable examples of Kane-Mele (KM) insulators \cite{RevModPhys.82.3045,molle2017buckled,liu2011quantum}. Due to
the buckling-type structure, the electronic and topological properties of these materials can be tuned via the application of a circularly off-resonant
polarized optical field as well as static external electric fields \cite{ezawa2012spin,liu2011quantum}. The tunable band gap makes them promising candidates for quantum-based technologies including topological spintronics and valleytronics \cite{xu2013large,schaibley2016valleytronics}. There is one challenge though: weak SOC in buckled X-enes limits their practical applications. 

In 2014, a new class of 2D materials known as transition-metal dichalcogenides (TMDCs) were discovered that exhibited large SOC \cite{zhang2014direct,ye2014probing}. The concurrence of spintronic and
valleytronic effects in these 2D semiconductors subsequently opened a promising route to investigate emergent electronic and photonic phenomena \cite{mak2016photonics,xu2014spin}. Within this larger family of materials, monolayer 1T$\mathrm{'}$-WTe$_{2}$ is an emerging class of materials that exhibits significant SOC required for the QSHI phase. Although the $\mathrm{1T'}$ phase is a rare composition, it has recently attracted huge interest due to the existence of topological phases \cite{qian2014quantum,wu2018observation}. 

From an application point of view, however,
2D M-Xenes are challenging due to their structural metastability and rapid oxidization in air \cite{peng2017observation}. An exception is a weak 3D stacked topological insulator $\mathrm{Bi_{14}Rh_{3}I_{9}}$ which is stable and structurally well-defined, though due to its structural complexity, it is not clear if it is possible to isolate a single layer \cite{rasche2013crystal}. The synthesis of a 2D material with concomitant topological
phases and significant SOC which overcomes the hindrances mentioned above is therefore highly desirable. Recently, jacutingaite ($\mathrm{Pt_{2}HgSe_{3}}$), a species belonging to the platinum group, has attracted attention
due to its significant band gap and topological nature \cite{marrazzo2018prediction}. $\mathrm{Pt_{2}HgSe_{3}}$ is a naturally occurring mineral, which was discovered in Brazil in 2008 \cite{cabral2008platinum}. Jacutingaite has the general
formula $\mathrm{Pt_{2}AX_{3}}$  ($\mathrm{A = Hg}$, $\mathrm{Cd}$ and $\mathrm{X = S, Se}$) and
is composed of layers coupled by the van der Waals interaction
in an $\mathrm{AA}$ configuration \cite{vymazalova2012jacutingaite,longuinhos2020raman}. Its properties are analogous to silicene and graphene. For example, it possesses a sandwich-like
structure consisting of a $\mathrm{Pt}$ layer sandwiched between two layers of $\mathrm{Se}$ and $\mathrm{Hg}$ ordered in a honeycomb structure with large SOC \cite{kandrai2020signature}. The Hg atoms in $\mathrm{Pt_{2}HgSe_{3}}$ form a buckled honeycomb lattice, similar to the structure found in 2D-Xenes \cite{molle2017buckled,liu2011quantum}.  The buckled honeycomb lattice structure and strong SOC are responsible for the topological properties of $\mathrm{Pt_{2}HgSe_{3}}$. The jacutingaite family has been investigated in various forms and configurations, including bulk and free-standing. For example, two members $\mathrm{Pt_{2}HgSe_{3}}$ and $\mathrm{Pd_{2}HgSe_{3}}$ of jacutingaite family have been experimentally synthesized in bulk form \cite{vymazalova2012jacutingaite,facio2019dual}. Both members are topological insulators with robust topological phases and ambient stability \cite{longuinhos2021mechanical,rademaker2021gate}. 

In the few-layer limit and for an odd number of layers,  $\mathrm{Pt_{2}HgSe_{3}}$ exhibits a QSHI state \cite{marrazzo2018prediction},  whereas in its bulk form it becomes semimetallic
and dual topological \cite{facio2019dual}. To observe the QSHI and valley-polarized quantum anomalous Hall effect (QAHE), the jacutingaite family has been explored in the free-standing form with inversion symmetry as well as on magnetic substrates, respectively \cite{facio2019dual,marrazzo2018prediction,rehman2022valley}. Through scanning tunneling microscopy (STM) and angle-resolved photoemission spectroscopy measurements, the Kane-Mele phase in monolayer $\mathrm{Pt_{2}HgSe_{3}}$ and dual in bulk $\mathrm{Pt_{2}HgSe_{3}}$ have been visualized and probed \cite{kandrai2020signature,cucchi2020bulk}. 

The low-energy physics of monolayer (ML)-jacutingaite near the Fermi level can be well-described by the KM model, which was initially developed for graphene \cite{RevModPhys.82.3045,molle2017buckled}. However, ML-jacutingaite has a much stronger SOC ($\approx0.15$ eV from the DFT and $ \approx0.5$ eV from many-body calculations) than graphene, leading to the opening of a significant band gap at the Dirac point \cite{marrazzo2018prediction,bafekry2020graphene}. Additionally, band gaps in 2D Xenes and M-Xenes have been modulated through various kinds of external stimuli,  for example,  electric, antiferromagnetic exchange, and off-resonance
circularly polarized optical fields \cite{hajati2016valley,wang2016spin,ezawa2013photoinduced, mak2012control,feng2019engineering,hao2020switch}. Recently, V. Vargiamidis \emph{et. al}, investigated topological phases of ML-jacutingaite in the presence of magnetic
exchange interactions and staggered sublattice potential \cite{vargiamidis2022tunable}. Their findings showed that when Rashba interaction is present, the system can exhibit both the QAHE and the quantum valley Hall effect (QVHE) simultaneously, under certain parameter regimes. Additionally, they investigated the DC Hall conductivities in different topological phases. In a similar study, M. Alipourzadeh \emph{et. al} investigated various topological phases of ML-jacutingaite under off-resonance optical field and staggered sublattice potential \cite{alipourzadeh2023photoinduced}. Furthermore, in the DC limit, they demonstrated that these distinct topological phases can be probed via the
spin- and valley-Hall conductivities at zero temperature. By exploiting the inversion symmetry in the presence of staggered sublattice potential in ML-jacutingaite, M. U. Rehman \emph{et.~al} uncovered several promising valley spin-based phenomena \cite{rehman2022jacutingaite}. More interestingly, they demonstrated that by reversing the staggered electric field from positive to negative induced spin splittings, Berry curvature (BC), and spin texture between two valleys can be swapped without destroying the $\mathrm{Z_2}$ topological hallmarks. However, the spin and valley resolved optical conductivities in finite frequency limits in different topological phases of ML-jacutingaites under staggered sublattice potential and off-resonance optical field still deserve rigorous exploration. We believe the current work fills some of the gaps in the understanding of this novel topological material.

In this paper, we systematically consider the Kane-Mele Hamiltonian in the presence of a sublattice staggered potential and off-resonance circularly polarized optical field. We examine the band structure of ML-jacutingaites in different topological regimes. Using the energy dispersion of ML-jacutingaites, we obtain the electronic density of states in these distinct topological phases. We go on to calculate the intrinsic spin and valley-resolved optical conductivities using analytical approaches of low-energy Hamiltonian and Kubo formalism. While the DC spin and valley Hall effects in jacutingaites have been explored theoretically \cite{vargiamidis2022tunable,alipourzadeh2023photoinduced}, here, we examine the finite-frequency spin and valley resolved optical response as a function of photonic energy and the impact of modulating the sublattice staggered potential and off-resonance circularly polarized optical field, providing ideas for subsequent experimental investigation. Furthermore, we demonstrate how the direction of the transverse spin and valley current can be controlled by energy of the incident light, and also analyze the impact of doping. All of these properties conclusively depend on the underlying topological manifold, which makes this work, we believe, interesting.

Our paper is organized as follows. We spell out the Kane-Mele (KM) Hamiltonian subject to circularly polarized optical and electrostatic fields and calculate the eigenenergies of the system in distinct topological phases in Sec.~\ref{A1}. We discuss the electronic density of states in Sec~\ref{B1}. In Sec~\ref{C1}, we analytically calculate the intrinsic spin and valley resolved longitudinal and Hall components of the conductivity of ML-jacutingaite materials while probing the effects of electric and optical fields.  In Sec.~\ref{E1}, we summarize and conclude. 

\begin{figure*}[ht!]
	\centering
	\includegraphics[width=0.8\linewidth]{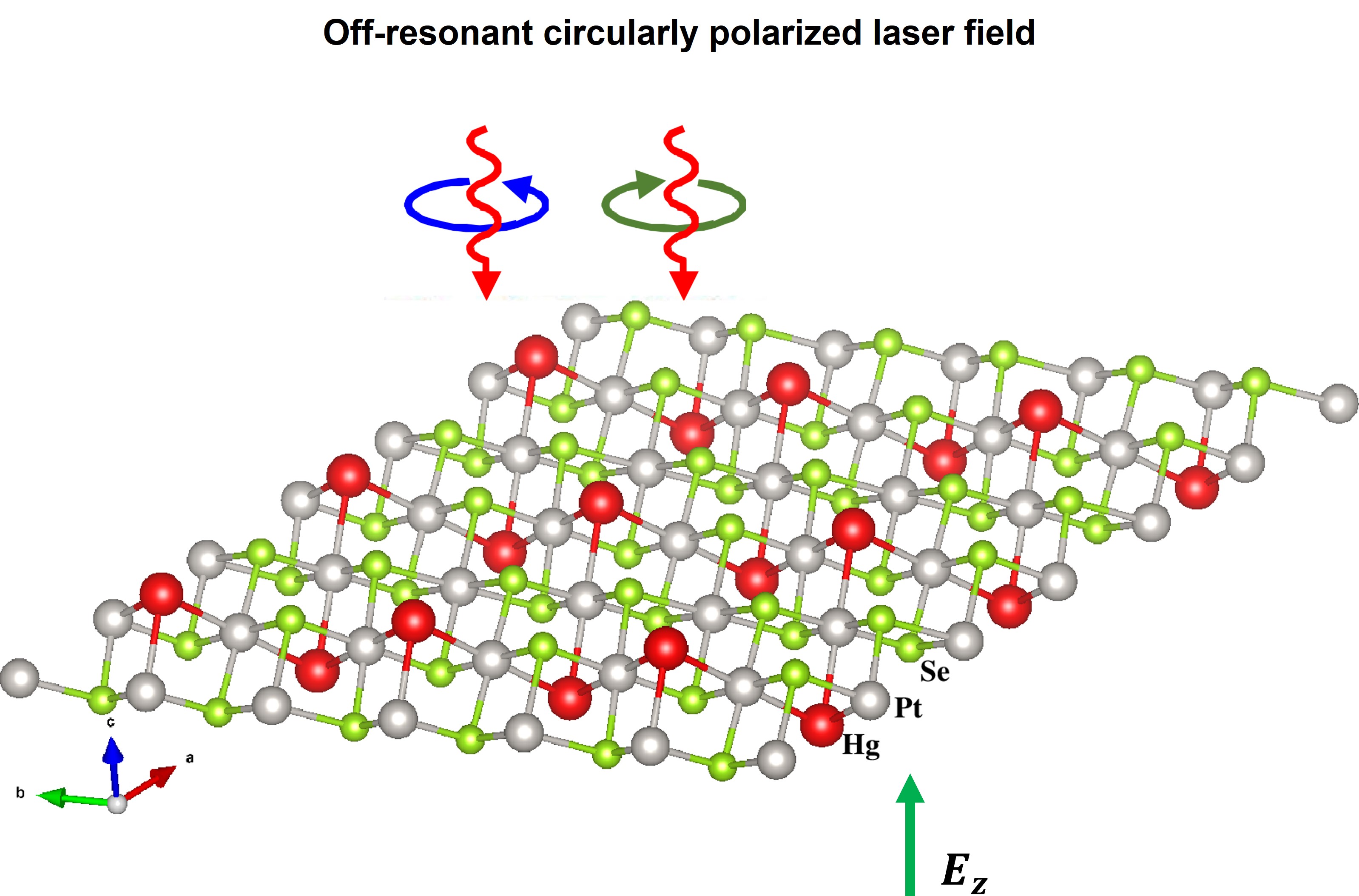}
	\caption{Schematic of ML-jacutingaite subjected to an external electric and off-resonant circularly polarized optical fields.}
	\label{fig1}
\end{figure*}


\section{System Hamiltonian and topological phases}\label{A1}
In the presence of an external perpendicular electrical field, the effective Hamiltonian of ML-jacutingaite at $K$ and $K'$ valleys can be written as \cite{ezawa2013photoinduced,vargiamidis2022tunable,alipourzadeh2023photoinduced}
\begin{equation}\label{a1}
	\hat{H}_{\eta, s}=\hbar v_{\mathrm{F}}\left(\eta k_x \sigma_x+k_y \sigma_y\right)+(\eta s \lambda_{so}+ \lambda_{z})\sigma_z,
\end{equation}
where, $k_{x}$ and $k_y$ are the components of crystal momentums and $v_{F}$ m/s is the Fermi velocity
in ML-jacutingaite. The parameter $\eta=\pm 1$ corresponds to the valley index ($K$ and $K^{'}$) in momentum space, $\sigma_{i}$ are the Pauli matrices in the sublattice space and $s=\pm1$ refers to the electron's spin. The first term in Eq.~(\ref{a1}) is the massless graphene-like
Hamiltonian. The second term in the Hamiltonian captures the Kane-Mele spin-orbit coupling (SOC) with $\lambda_{so}$= 81.2 meV \cite{marrazzo2018prediction}. ML-jacutingaite ($\mathrm{Pt_{2}HgSe_{3}}$) is a quantum spin Hall insulator (QSHI) with large band gaps, with an SOC that is about twenty times larger than in silicene \cite{ezawa2013spin}. The third term represents the perpendicular electric field $E_{z}$ creating the potential $\lambda_{z}$ and is responsible for breaking the $A$, $B$ sublattice inversion symmetry. 

In order to incorporate circularly polarized light in our system we use the formalism developed by Floquet which treats light as a time-periodic external electromagnetic perturbation. Consider that circularly polarized light irradiated onto ML-jacutingaite then the vector potential $\boldmath{A}(t)$, takes the form
\begin{equation}\label{a7}
	\text{\boldmath{A}}(t)=A_{0}\big(\gamma \sin \omega_{0} t, \cos \omega_{0} t ),
\end{equation}
where $\gamma=\pm 1$ represents the helicity of right and left circularly polarized light, $A_{0}=E_{0}/\omega_{0}$, $E_{0}$ is the amplitude of the electromagnetic field, and$\omega_{0}$ is the frequency of the incident light. The incident intensity is characterized~\cite{ezawa2013photoinduced} by the parameter $\mathcal{A}^2=\lvert e a E_{0}/\hbar\omega_{0}\rvert^2$ where $a$ is the lattice constant. The periodicity of the perturbation is reflected in the effective Hamiltonian \cite{zubair2022valley}
\begin{equation}\label{a6aa}
	\hat{H}_{\eta, s}(t)=\hat{H}_{\eta, s}^{0}+\hat{V}(t),
\end{equation}
where the unperturbed part is identical to the Hamiltonian~\eqref{a1} and the time-dependent periodic part can be written as 
\begin{equation}\label{a2b}
	\hat{V}(t)= \frac{e v_{F}}{\hbar}( \eta A_{x}(t)\sigma_{x} + A_{y}(t)\sigma_{y}).
\end{equation}
For incident frequencies much larger than Fermi energy, i.e., $a\omega_{0}\gg v_F$ and low optical intensities $e a E_{0} / \hbar \omega_{0} = \mathcal{A}\ll 1$, the periodic term in Eq.~\eqref{a6aa} can be calculated~\cite{kitagawa2011transport}:
\begin{equation}\label{a8}
	\hat{V}(t)=
 \frac{\big[H_{-1}, H_{1}\big]}{\hbar\omega_{0}}+\mathcal{O}(\omega_{0}^{-2}),
\end{equation}
where $H_{\pm1}$ represent the Fourier components of the Hamiltonian and are given by
\begin{equation}\label{a9}
	\hat{H}_{\pm}=\frac{\omega_{0}}{2\pi}\int_{0}^{\frac{2\pi}{\omega_{0}}}dt\,\hat{H}_{0}(t)e^{\pm i\omega_{0} t}.
\end{equation}
These conditions are readily met in typical scenarios and common optical frequencies and powers. For example, for typical values for a $100~$mW radiation of wavelength $500~$nm, resulting in a frequency of $0.6~$THz, a lattice constant $a\approx 3~$\AA, Fermi velocity in the range of $10^5~$m/s, we have $a\omega_0/v_F\approx 11$ and $\mathcal{A}\approx 7\times 10^{-7}$. For THz radiation, this condition is even more strictly satisfied. Under these conditions~\cite{rodriguez2017casimir}, we can calculate the perturbative term, 
\begin{equation}\label{a12}
\lambda_\omega=\frac{\big[H_{-1}, H_{1}\big]}{\hbar\omega_{0}}=\frac{\gamma (e A_{0}v_{F})^2}{\hbar\omega_{0}}.
\end{equation}
Given that $A_0\propto \omega_0^{-1}$, we observe that $\lambda_{\omega}\propto \omega_{0}^{-3}$. All in all, the periodic perturbation results in the effective Hamiltonian,
\begin{equation}\label{b1a}
	\hat{H}_{\eta, \mathrm{s}}=\hbar v_{\mathrm{F}}\left(\eta k_x \sigma_x+k_y \sigma_y\right)+(\eta s \lambda_{so}  + \lambda_{z} + \eta\lambda_\omega)\sigma_z.
\end{equation}
It should be noted that the light field of frequency $\omega_0$ induces additional coupling between the energy bands \cite{oka2009photovoltaic}, leading to the formation of gaps in the band structure.  These gaps typically occur at energies around $n \hbar \omega_0 / 2$ ( $n= \pm 1, \pm 2$, etc).
The energy eigenvalues for the Hamiltonian are
\begin{equation}\label{b1}
	E_{t}^{\eta, s}=t \sqrt{\left(\hbar v_{\mathrm{F}} k\right)^2+\left(\Delta_{\eta, s}\right)^2},
\end{equation}
where $t= \pm$ stands for the electron and hole bands and $\Delta_{\eta, s}=(\eta s \lambda_{so}  + \lambda_{z} + \eta\lambda_{\omega})$ is the Dirac mass responsible for the effective gap.  The Hamiltonian in Eq.~\eqref{b1a} is valid as long as $\left|E_{t}^{\eta, s}\right|<\hbar \omega_0 / 2$.

Note that the presence of the off-resonant circularly polarized optical field $\lambda_{\omega_0}$ and the sublattice staggered
potential $\lambda_{z}$ break the TRS and inversion symmetries leading to nonzero Berry curvature (BC) and hence nonzero anomalous Hall conductivity. The Berry curvature in the out-of-plane
direction for the $t$'th band comes out as \cite{shah2022optical}
\begin{equation}\label{b2}
	\Omega_{\eta, \mathrm{s}}^t(k)=-2 \hbar^2 \operatorname{Im} \sum_{\mathrm{t} \neq \mathrm{t}^{\prime}} f_{t}^{\eta, s}(k)  \frac{\langle v_{x}\rangle \langle v_{y}\rangle}{\left(E_{t}^{\eta, s}-E_{t'}^{\eta, s}\right)^2},
\end{equation}
where $\langle v_{x}\rangle \langle v_{y}\rangle=\left\langle\psi_{t}^{\eta, s}\left|v_{\mathrm{x}}\right| \psi_{ to}^{\eta, s}\right\rangle\left\langle \psi_{ to}^{\eta, s}\left|v_{\mathrm{y}}\right| \psi_{t}^{\eta, s}\right\rangle$, $f_{t}^{\eta, s}(k)=\left[1+\exp \left(E_{t}^{\eta, s}-\mu_{\mathrm{f}}\right) / k_{\mathrm{B}} T\right]^{-1}$ is the Fermi Dirac distribution function for the $t$'th band with chemical potential $\mu_{F}$, and $\psi_{t}^{\eta, s}$ is the Bloch
state with energy eigenvalues $E_{t}^{\eta, s}$. 

The interplay between the sublattice staggered
potential $\lambda_{z}$ and laser $\lambda_{\omega_0}$ field in ML-jacutingaite offers the possibility of attaining various steady-state phases, discernible from the effective Dirac mass $\Delta_{\eta, s}$. In the insulating phase when the chemical potential level lies in the bulk gap, the spin-valley-dependent Chern number can be expressed as
\begin{equation}\label{b4}
	\mathcal{C}_{\eta, s}=\frac{1}{2 \pi} \int d^2 k\,\, \Omega_{\eta, s}^t(k),
\end{equation}
which can be written in simplified form as
\begin{equation}\label{b5}
	\mathcal{C}_{\eta, \mathrm{s}}=\frac{\eta}{2} \operatorname{sgn}\left(\Delta_{\eta, \mathrm{s}}\right)
\end{equation}
wherein $\operatorname{sgn}(x)$ is the sign function. From Eq. \eqref{b5}, we can characterize the topological insulator state by the charge Chern number and spin Chern number.  The spin and valley Chern numbers can be calculated at each
point as
\begin{equation}\label{b6}
	\mathcal{C}_{\mathrm{s}}=\frac{\left(\mathcal{C}_{\uparrow}-\mathcal{C}_{\downarrow}\right)} {2},
\end{equation}
and 
\begin{equation}\label{b7}
	\mathcal{C}_{\mathrm{v}}=\frac{\left(\mathcal{C}_{\mathrm{K}}-\mathcal{C}_{\mathrm{K}^{\prime}}\right)}{2},
\end{equation}
where $\mathcal{C}_{\uparrow(\downarrow)}=\sum_\eta C_{\eta, \uparrow(\downarrow)}$ and $\mathcal{C}_{\mathrm{K}\left(\mathrm{K}^{\prime}\right)}=\sum_{\mathrm{s}} \mathcal{C}_{\mathrm{K}, \mathrm{s}\left(\mathrm{~K}^{\prime}, \mathrm{s}\right)}$. Chern numbers determine the topological properties, with $\mathcal{C}\neq 0$ indicating a nontrivial topological phase. In this case, the sign of Dirac mass $\Delta_{\eta, s}$ determines the topological phase of the material. A topological phase transition takes place when the sign of the effective
Dirac mass $\Delta_{\eta, s}$
switches. The boundaries between phases are defined by $\Delta_{\eta, s}$= 0 \cite{ezawa2012topological,ezawa2013photoinduced}.

\begin{figure*}[t!]
	\centering
	\includegraphics[width=0.6\linewidth]{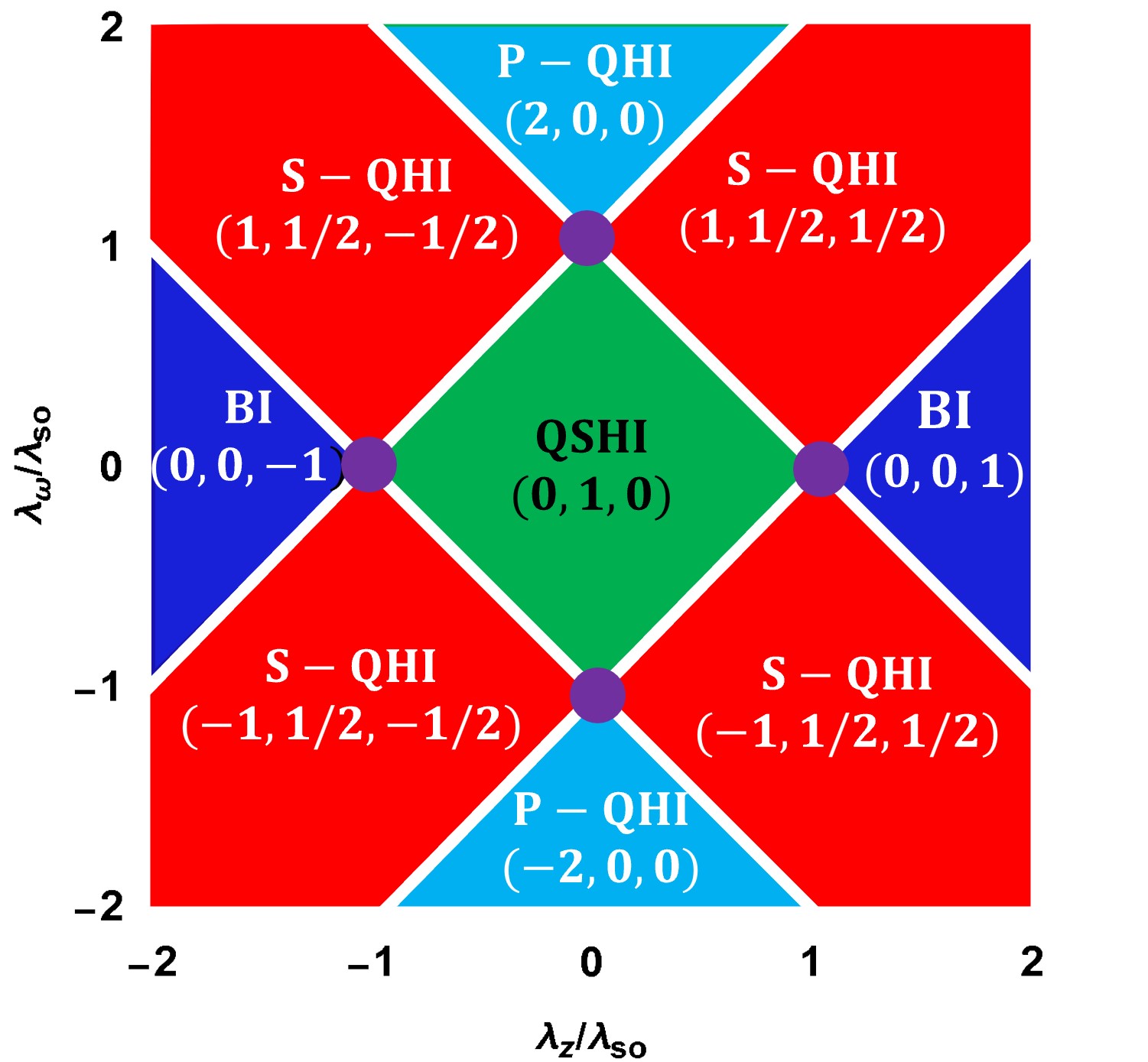}
	\caption{The phase diagram of the monolayer of
		jacutingaite as a function of $\lambda_{z}/\lambda_{so}$ and $\lambda_{\omega}/\lambda_{so}$. The distinct electronic phases are labeled by different colors and are indexed by the total, spin, and valley Chern numbers ($\mathcal{C}$, $\mathcal{C}_{s}$ and $\mathcal{C}_{v}$).}
	\label{phase}
\end{figure*}

\begin{figure*}[ht!]
	\centering
	\includegraphics[width=0.45\linewidth]{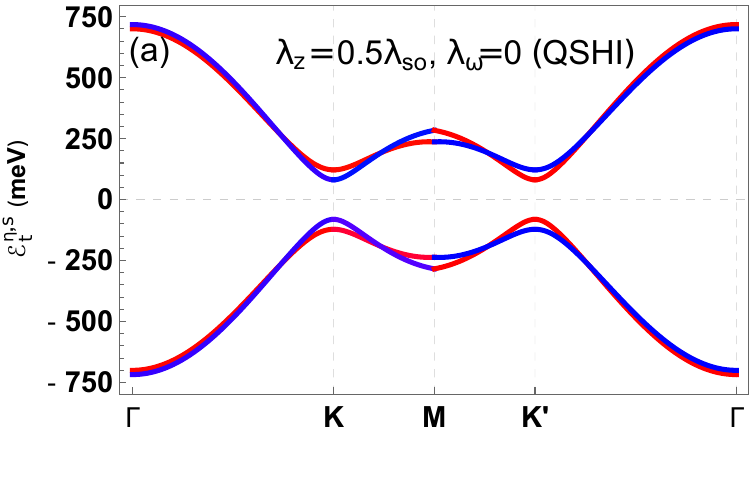}
	\includegraphics[width=0.450\linewidth]{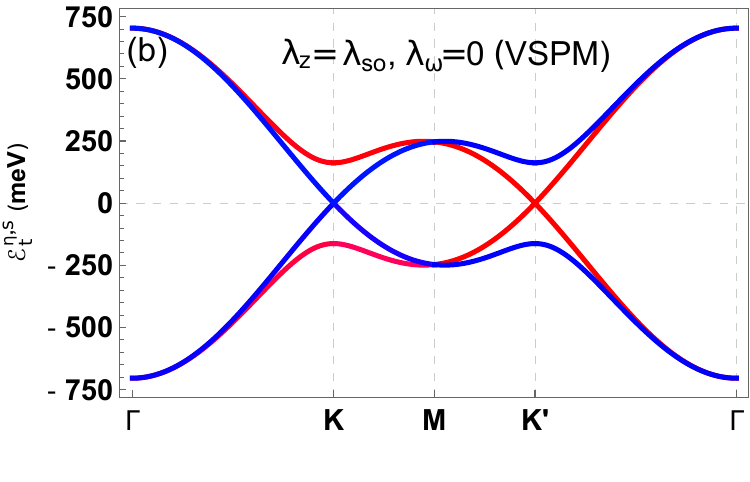}\\
	\includegraphics[width=0.450\linewidth]{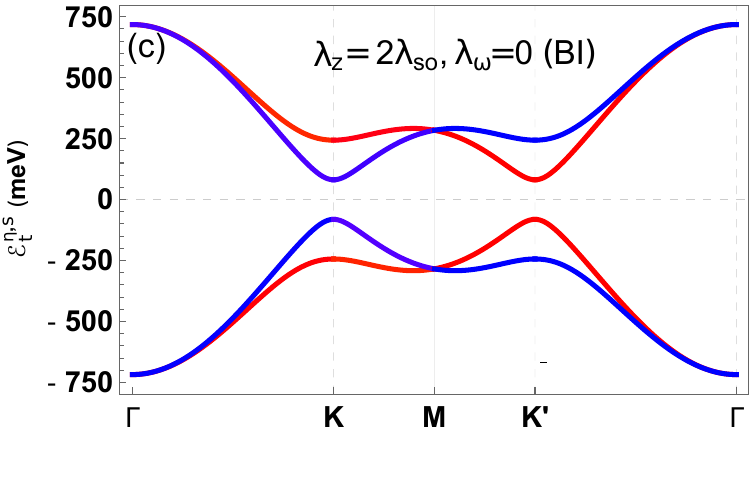}
	\includegraphics[width=0.450\linewidth]{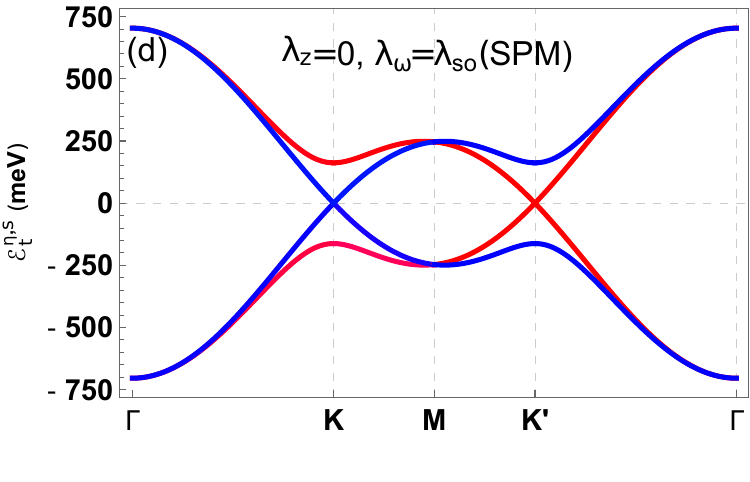}\\
	\includegraphics[width=0.450\linewidth]{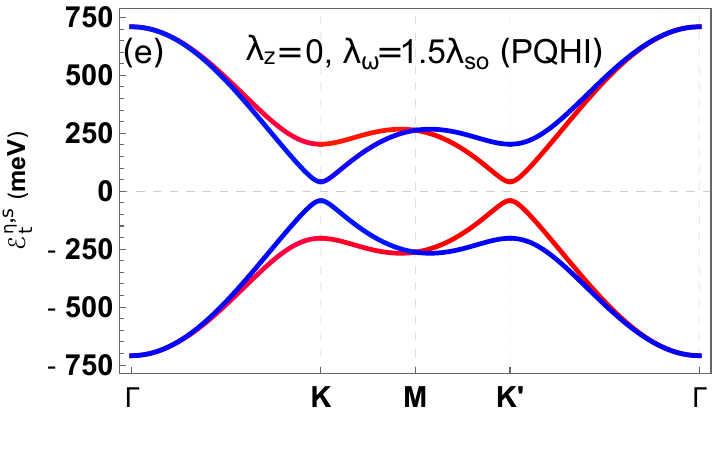}
	\includegraphics[width=0.450\linewidth]{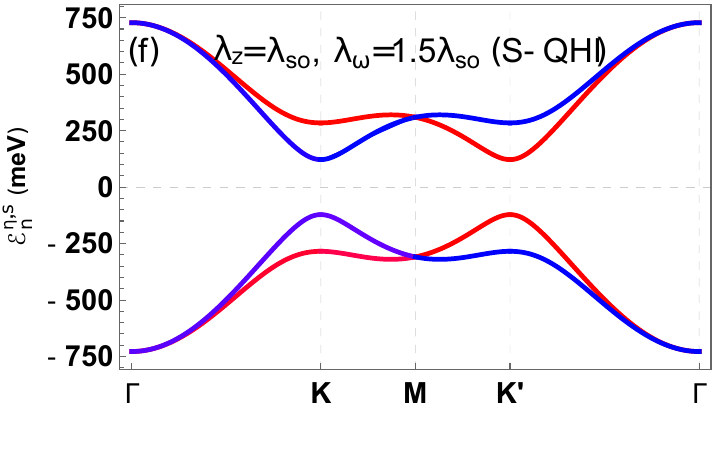}
	\caption{Band structure of jacutingaite in distinct topological phases at the $K$ valley. (a) QSHI ($\lambda_{z}=0.5\lambda_{so}$, $\lambda_{\omega}=0$), (b) VSPM  ($\lambda_{z}=\lambda_{so}$, $\lambda_{\omega}=0$), (c) BI ($\lambda_{z}=1.5\lambda_{so}$,  $\lambda_{\omega}=0$), (d) SPM ($\lambda_{z}=0$,  $\lambda_{\omega}=\lambda_{so}$), (e) P-QHI ($\lambda_{z}=0$, $\lambda_{\omega}=1.5\lambda_{so}$) and (f) S-QHI ($\lambda_{z}=\lambda_{so}$,  $\lambda_{\omega}=1.5\lambda_{so}$) respectively. The blue and red curves refer to spin-up and spin-down energy bands, respectively.}
	\label{Bandgaps}
\end{figure*}

Figure~\ref{phase} summarizes the phase diagram for ML-jacutingaite, showcasing distinct topological phases separated by phase transition boundaries in the $\lambda_{z}/\lambda_{so}$ and $\lambda_{\omega}/\lambda_{so}$ plane. The total, spin, and valley Chern numbers $\left(\mathcal{C}, \mathcal{C}_{\mathrm{s}}\right.$, $\mathcal{C}_{\mathrm{v}}$) identify different topological phases (which are indicated by different colors). For each of these topological phases, we present typical band structures
of ML-jacutingaite in Figs.~\ref{Bandgaps}(a)-(f). The energy spectrum in Figs.~\ref{Bandgaps}(a)-(f) corresponds to different color regions in Fig. \ref{phase}. The external stimuli $\lambda_{z}$ and $\lambda_{\omega}$ affect the effective gap and can be used to control the Dirac mass $\Delta_{\eta, s}$ for the different Dirac cones. By modulating the external fields we reveal a rich variety of electronic phases in monolayer jacutingaite associated with the Hall effect. 

Depending on how many Dirac cones have non-zero mass gaps and non-zero Chern numbers, a galaxy of quantum phases, for example, quantum spin Hall insulator (QSHI), spin-polarized
metal (SPM), spin-polarized quantum Hall insulator (S-QHI), spin valley polarized metal (SVPM), normal or band insulator
(BI), single Dirac cone (SDC), and polarized
spin quantum Hall insulator (P-QHI) become possible. This is the rich interplay of various terms in the system Hamiltonian. The classification of phases is described in Table~\ref{mytab}.  In the unperturbed case, when $\lambda_{\omega}=\lambda_{z}=0$, the bands are spin degenerate and are separated by an insulating gap of 2$\lambda_{so}$,
which makes it a QSHI phase characterized by $\mathcal{C}_{\uparrow}=1$ and $\mathcal{C}_{\downarrow}=-1$, resulting in $\mathcal{C}_{\mathrm{s}}=1$, while $\mathcal{C}=\mathcal{C}_{\mathrm{v}}=0$. This is in agreement with the results predicted in Ref \cite{vargiamidis2022tunable}. Initially, we fix the off-resonant irradiated optical field $\lambda_{\omega}$=0 and tune the staggered sublattice potential $\lambda_{z}$.  As long as $\lambda_{z}<\lambda_{so}$,
the ML-jacutingaite remains inside the QSHI phase with a spin splitting and two energy gaps as illustrated in Fig.~\ref{Bandgaps}(a).  However, it could be noted that when $\lambda_{z}=\lambda_{so}$, the spin-down effective gap $\left|\lambda_{z}-\eta \lambda_{so}\right|$ closes, reflecting the graphene-like nature of the VSPM. The corresponding
band structure of the VSPM state for the $K$ valley is shown in Fig.~\ref{Bandgaps}(b). Further tuning  $\lambda_{z}$ leads to a quantum phase transition from the QSHI to BI and the gaps are reopened (see in Fig.~\ref{Bandgaps}(c)). In the BI regime, the spin Chern number $\mathcal{C}_{\mathrm{s}}=0$ while the valley is $\mathcal{C}_{\mathrm{v}}=1$. The BI regime can be related to the $\mathrm{QVHI}$ state. 
Conversely, if we increase $\lambda_{\omega}$ while keeping $\lambda_{z}=0$, the spin-down gap closes at $\lambda_{\omega}=\lambda_{so}$, 
where ML-jacutingaite is a semimetal (Fig.~\ref{Bandgaps}(d)). The spin-polarized metal (SPM) appears at this point. Further increasing $\lambda_{\omega}$ results in the gap reopening and the system enters into quantum anomalous Hall effect regime triggered by light, namely a P-QHI, as depicted in Fig.~\ref{Bandgaps}(e). The associated Chern numbers in this phase are $\mathcal{C}=2$ and $\mathcal{C}_{\mathrm{s}}=\mathcal{C}_{\mathrm{v}}=0$.  What will happen when both the staggered electric potential and off-resonant laser fields are turned on (i.e., $\left| \pm \lambda_{z} \pm \lambda_{\omega}\right|>\lambda_{so}$)? When both of the fields are switched on, we can observe
a phase transition from the AQHI to an electromagnetically induced $\mathrm{S}$-$\mathrm{QHI}$ phase. In the $\mathrm{S}$-$\mathrm{QHI}$ regime, the spin- and valley-resolved Chern numbers become $\mathcal{C}_{\uparrow}=1$, $\mathcal{C}_{\downarrow}=0,  
\mathcal{C}_{\mathrm{K}}=1$, and $\mathcal{C}_{\mathrm{K}^{\prime}}=0$. The band structure in the $\mathrm{S}$-$\mathrm{QHI}$ phase is shown in Fig.~\ref{Bandgaps}(f). 

\begin{table}[t]
	{\caption{Spin- and valley-resolved and spin (valley) Chern numbers for different topological phases in ML-jacutingaites. \label{mytab} }}
	\centering
	\begin{tabular}{lrrrrrllc}
		\hline Topological phase & $\mathcal{C}_{\uparrow}$ & $\mathcal{C}_{\downarrow}$ & $\mathcal{C}_{\mathrm{s}}$ & $\mathcal{C}_{\mathrm{K}}$ & $\mathcal{C}_{\mathrm{K'}}$ & $\mathcal{C}_{\mathrm{v}}$ &  \\
		\hline QSHI & 1 & -1 & 1 & 0 & 0 & 0   \\
		BI (QVHI) &  0 & 0 & 0 & 1 & -1 & 1   \\
		QSHI & 1 & -1 & 1 & 0 & 0 & 0 \\
		P-QHI & 1 & 1 & 0 & 1 & 1 & 0  \\
		S-QHI & 1 & 0 & 1 & 1 & 0 & 1  \\
		\hline
	\end{tabular}
\end{table}

\begin{figure*}[ht!]
	\centering
	\includegraphics[width=0.45\linewidth]{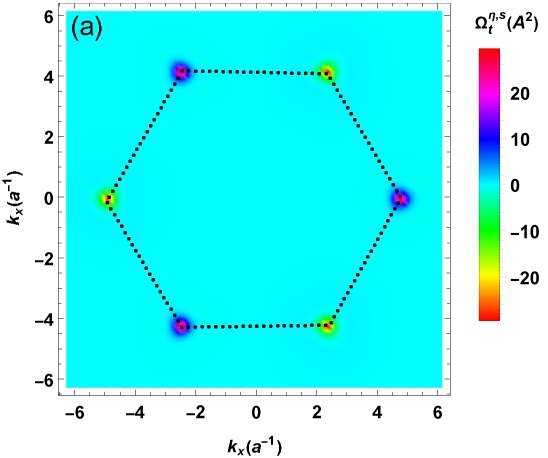}
	\includegraphics[width=0.450\linewidth]{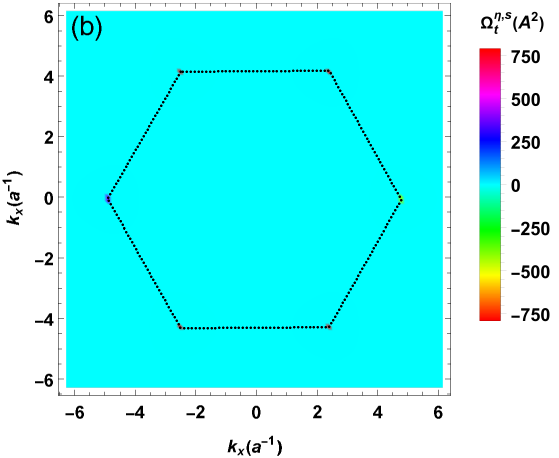}\\
	\includegraphics[width=0.450\linewidth]{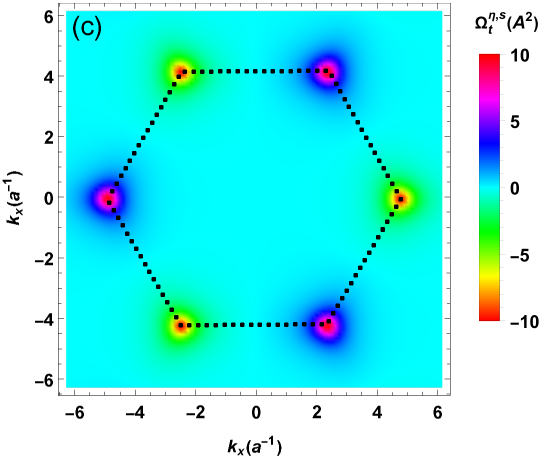}
	\includegraphics[width=0.450\linewidth]{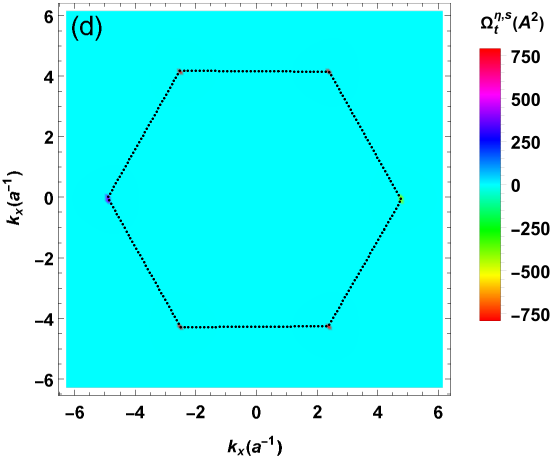}\\
	\includegraphics[width=0.450\linewidth]{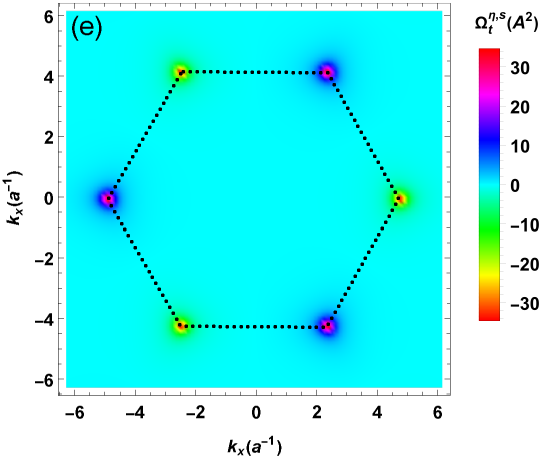}
	\includegraphics[width=0.450\linewidth]{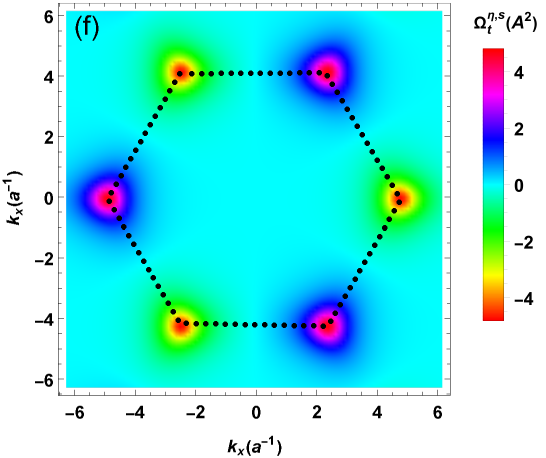}
	\caption{Berry curvature of jacutingaite in distinct topological phases at the $K$ and $K'$ valleys. (a) QSHI ($\lambda_{z}=0.5\lambda_{so}$, $\lambda_{\omega}=0$), (b) VSPM  ($\lambda_{z}=\lambda_{so}$, $\lambda_{\omega}=0$), (c) BI ($\lambda_{z}=1.5\lambda_{so}$,  $\lambda_{\omega}=0$), (d) SPM ($\lambda_{z}=0$,  $\lambda_{\omega}=\lambda_{so}$), (e) P-QHI ($\lambda_{z}=0$, $\lambda_{\omega}=1.5\lambda_{so}$) and (f) S-QHI ($\lambda_{z}=\lambda_{so}$,  $\lambda_{\omega}=1.5\lambda_{so}$) respectively. }
	\label{BC}
\end{figure*}
It is observed that when $\lambda_{z}<\lambda_{so}$ and $\lambda_{\omega}=0$, the spin-up and spin-down band gaps remain open and support the QSHI regime. In the QSHI phase, the spin-up Chern number $\mathcal{C}_{\uparrow}=1$ and spin-down Chern number $\mathcal{C}_{\downarrow}=-1$, resulting in $\mathcal{C}_{\mathrm{s}}=1$, while the valley Chern number $C=\mathcal{C}_{\mathrm{v}}=0$, leading to a QSHI state that originates from the SOC-induced band gap. This is clearly seen in Figs. \ref{Bandgaps}(a) and (b) for both the $K$ and $K'$ valleys. At the critical sublattice potential $\lambda_{z}=\lambda_{so}$ and $\lambda_{\omega}=0$, the gap closes for spin-up and spin-down electrons of the $K$ and $K'$ valleys, respectively. resulting in the VSPM state, as depicted in Figs. \ref{Bandgaps}(c) and (d) for the $K$ and $K'$ valleys. The VSPM state is shown in the phase diagram by a purple circle. This indicates that the structure behaves metallic at the critical field. The band structure is not the same at the $K$ and $K'$ valleys which is a signature of the aforementioned band inversion. When the sublattice potential further increases i.e. $\lambda_{z}>\lambda_{so}$ and $\lambda_{\omega}=0$, the gap reopens and signifies a phase transition from a QSHI phase to a normal insulator state as presented in Figs. \ref{Bandgaps}(e) and (f) for the $K$ and $K'$ valleys. The corresponding spin-up, spin-down, and valley Chern numbers for the BI (QVHI) phase are $\mathcal{C}_{\uparrow}=0$, $\mathcal{C}_{\downarrow}=0$ and $C=\mathcal{C}_{\mathrm{v}}=1$, respectively. Conversely, if we keep  $\lambda_{z}=0$ and $\lambda_{\omega}=\lambda_{so}$ then the system reaches an SPM state, as shown in Figs. \ref{Bandgaps}(g) and (h) for the $K$ and $K'$ valleys. In the SPM state, the band gaps are closed for the spin-up and spin-down states in the corresponding valleys. Increasing the optical field intensity beyond $\lambda_{so}$ reopens the
gap, and the system enters the P-QHI phase, as depicted in
Figs. \ref{Bandgaps}(i) and (j) for the $K$ and $K'$ valleys with spin-up, spin-down and valley Chern numbers $\mathcal{C}_{\uparrow}=1$, $\mathcal{C}_{\downarrow}=1$ and $C=\mathcal{C}_{\mathrm{v}}=0$, respectively. Finally, if we tune the optical and electric field intensity simultaneously, i.e $|\pm\lambda_{z}\pm\lambda_{\omega}=0|>\lambda_{so}$, all degeneracies are broken, leading to an S-QHI phase, as illustrated in Figs. \ref{Bandgaps}(k) and (l) for the $K$ and $K'$ valleys.  In the S-QHI phase, the associated spin-up, spin-down, and valley Chern numbers $\mathcal{C}_{\uparrow}=1$, $\mathcal{C}_{\downarrow}=0$ and $C=\mathcal{C}_{\mathrm{v}}=0$, respectively. 

\begin{figure*}[ht!]
	\centering
	\includegraphics[width=0.35\linewidth]{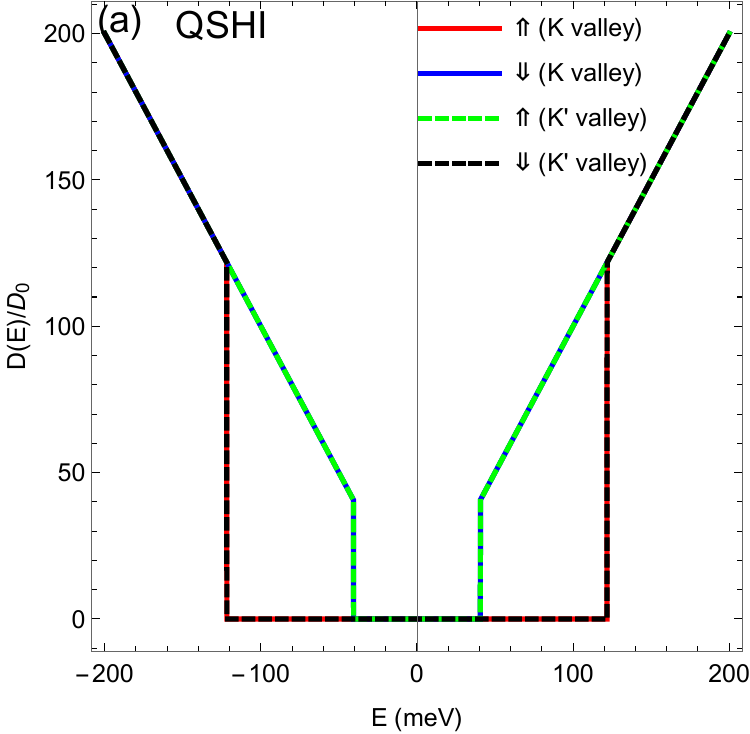}
	\includegraphics[width=0.35\linewidth]{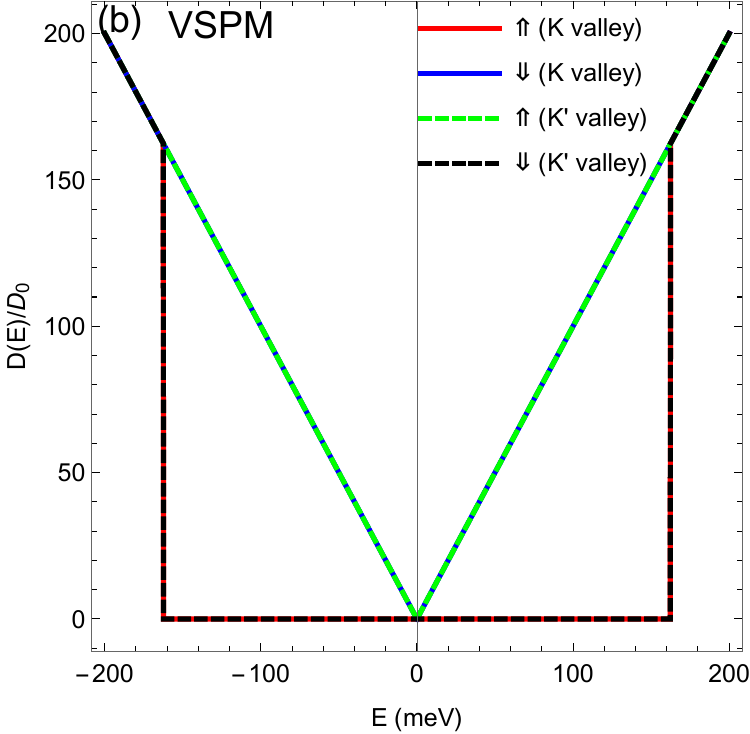}\\
	\includegraphics[width=0.35\linewidth]{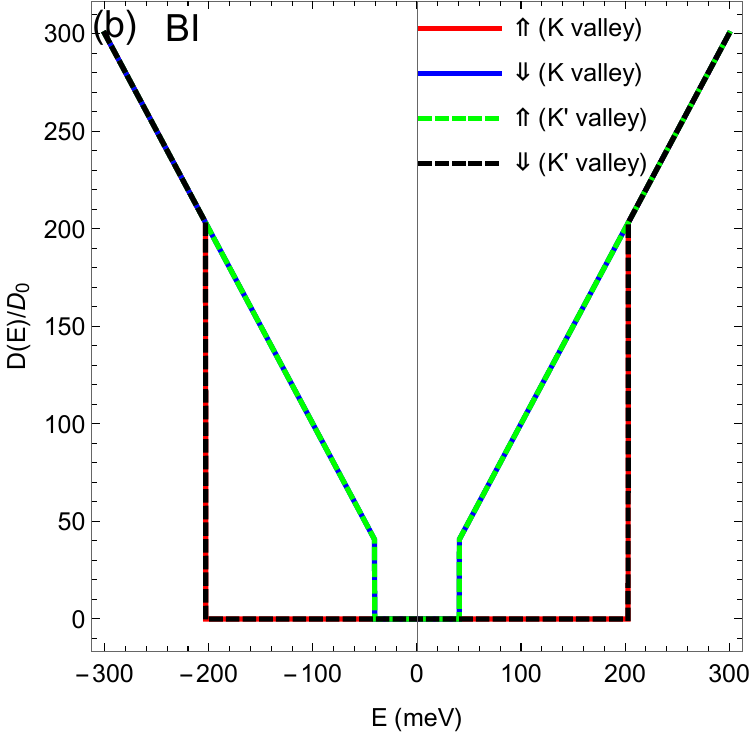}
	\includegraphics[width=0.35\linewidth]{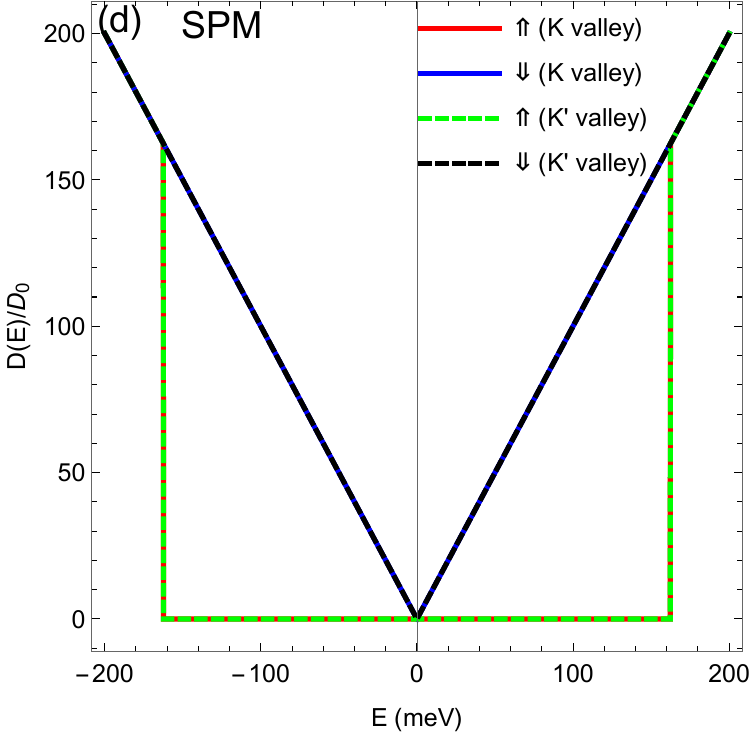}\\
	\includegraphics[width=0.35\linewidth]{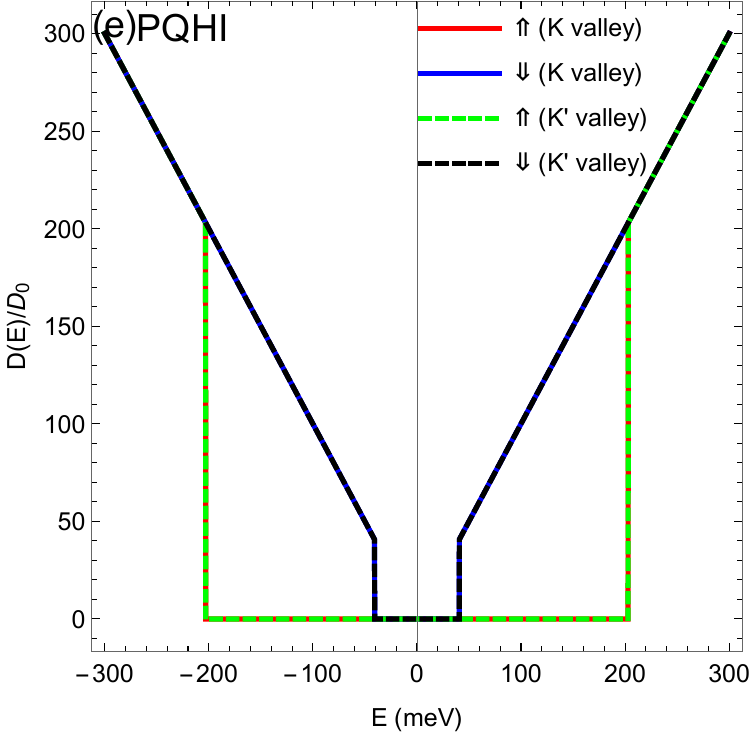}
	\includegraphics[width=0.35\linewidth]{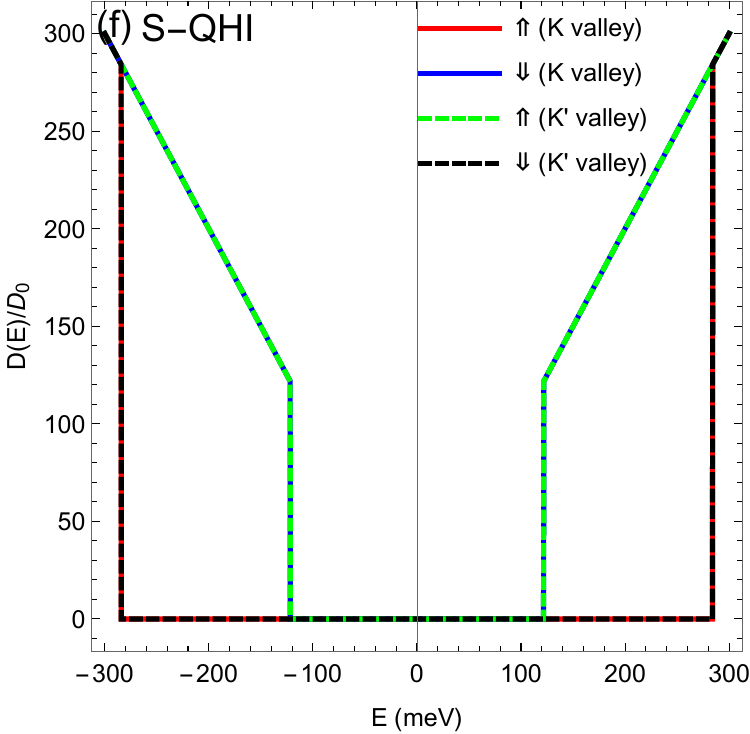}
	\caption{The low-energy electronic density of states of jacutingaite at the $K$ and $K'$ valleys in distinct topological phases as shown on the phase diagram. The solid red and blue curves refer to the spin-up and spin-down density of states in the $K$ valley respectively. Similarly, the dashed green and black curves correspond to the spin-up and spin-down density of states in the $K'$ valley respectively.}
	\label{DOS}
\end{figure*}

Figs.~\ref{BC}(a)–-(f) show the spin and valley-resolved Berry curvatures  of ML-jacutingaite in distinct topological phases. It has been shown that ML-jacutingaite exhibits a semimetallic character in the absence of spin-orbit coupling, where the conduction and valence bands meet \cite{rehman2022jacutingaite}. The spin-up and spin-down bands are degenerated due to the inversion symmetry in the layer. By applying an external staggered electric field to ML-jacutingaite, without considering spin-orbit coupling effects, the material is transforming from a semimetallic state to a quantum valley Hall insulator (QVHI) phase. Fig.~\ref{BC}(a) displays the valley contrasting Berry curvature when $\lambda_{z}=0.5\lambda_{so}$ and $\lambda_{\omega}=0$. We can see that the spin-up and spin-down bands are no longer overlapping indicating lifting of the spin degeneracy, which is a consequence of the combined effects of the staggered sublattice potential and the SOC. It must be noted that the staggered sublattice potential does not disrupt the underlying topological nature of the material, specifically, its $\mathrm{Z_2}$ topological number is still preserved. ML-jacutingaite is still a QSHI as it is before the application of staggered field. From the valley contrasting Berry curvature (opposite peaks for opposite valleys) it is obvious that the system is exhibiting QVHI phase. The simultaneous presence of the twinned QSHI and QVHI phases in ML-jacutingaite, despite the applied electric field, demonstrates a coupled spin-valley Hall effect or quantum spin-valley Hall effect. Next, we calculate the Berry curvature in the VSPM state. In this phase, the spin-down (spin-up) band closes at the $K$ $(K')$ valley [see Fig.~\ref{Bandgaps}(b)]. Sharp peaks can be seen at the $K$ and $K'$ valleys in Fig.~\ref{BC}(b). In the BI phase, the spin-up and spin-down bands are opening again in the respective valleys as presented in Fig.~\ref{Bandgaps}(c). In contrast to the QSHI phase, here the Berry curvature signs are different at the $K$ and $K'$ valleys [see Fig.~\ref{BC}(b)]. In the SPM state, we can observe similar trend as in the case of VSPM phase. Finally, in the P-QHI and S-QHI phases, the valley contrasting Berry curvatures are shown in Figs.~\ref{BC}(e) and (f) respectively.

\section{Density of states}\label{B1}
From Eq.~\eqref{b1}, we can obtain the electronic density of states (DOS) per unit area which is given by,
\begin{equation}\label{c1}
	D(E)=\sum_{\eta s} \sum_k \delta\left(E-E^{\eta, s}(\boldsymbol{k})\right) .
\end{equation}
Working in the continuum limit, we can express 
\begin{equation}\label{c2}
	D(E)=\sum_{\eta, s} D_{\eta, s}(E),
\end{equation}
where
\begin{equation}\label{c3}
	D_{\eta, s}(E)=D_{0}|E|\Theta\bigg(|E|-|\Delta_{\eta, s}|\bigg),
\end{equation}
where $\Theta$ is the Heaviside function. The $D(E)$ can be obtained as
\begin{equation}\label{c4}
	D(E)=D_0 \sum_{s, \eta}\Theta\left(|E|-\left|\Delta_{\eta, s}\right|\right)
\end{equation}
where $D_0=1 /\left(2 \pi v_{F}^2 \hbar^2\right)$. We have shown the dimensionless density of states $D(E) / D_0$ in Figs. \ref{DOS}(a)--(f) in different topological phases. In Fig.~\ref{DOS}(a), we have plotted $D(E) / D_0$ as a function of energy in the QSHI phase for both spins and valleys.  The two energy gaps of the spin-up and spin-down are clearly reflected in the density of states. In the QSHI phase, initially, the density of states is zero for energies below the lowest gap. A jump can be seen at the optical excitation energy of the spin-down (spin-up) electron at the $K$ ($K'$) valley. Similarly, the second jump is originates at the optical transition of the spin-up (spin-down) electron at the $K$ ($K'$) valley. After the jumps $D(E)/ D_0$ continues to increase linearly as shown in Fig.~\ref{DOS}(a). In the VSPM phase $\lambda_{z}=\lambda_{so}$ and $\lambda_{\omega}=0$, the lowest band gap of the spin-down (spin-up) at the $K$ ($K'$) valley closes. The $D(E) / D_0$ of the spin-down electron at the $K$ valley and spin-up electron at the $K'$ valley grows linearly from zero energy, 
while that of the spin-up electron at the $K$ valley and spin-down electron at the $K'$ valley are jumping vertically at $E=\lambda_{so}+\lambda_{z}$, and further increases linearly as depicted in Fig.~\ref{DOS}(b). As $\lambda_{z}=\lambda_{so}$ is increased further, the system transitions into the BI (QVHI) phase and the spin-down energy gap reopens. The two jumps in $D(E) / D_0$ reflect the energy gaps in the electronic structure of jacutingaite at the $K$ and $K'$ valleys as depicted in Fig.~\ref{DOS}(c). In the SPM phase ($\lambda_{\omega}=\lambda_{so}$ and $\lambda_{z}$=0), the band gaps for the spin-down electron in both valleys are closed as shown in Fig.~\ref{Bandgaps}(d). In this phase, $D(E) / D_0$ for the spin-down electron in both valleys increases
linearly out of zero energy as shown in Fig.~\ref{DOS}(d).  For the spin-up electron, a jump can be seen at a higher value
due to the availability of the upper-gapped band. As the optical field $\lambda_{\omega}$ strength increases, ML-jacutingaite transitions into the S-QHI phase (a distinct scenario compared to the previous BI (QVHI) phase), and the lowest energy gaps of the spin-down electron at the $K$ and $K'$ valley reopen. Again there are two jumps in $D(E) / D_0$ which show the two gaps for the spin-up and spin-down electrons at the $K$ and $K'$ valleys as shown in Fig.~\ref{DOS}(e). In the S-QHI regime, $D(E) / D_0$ are presented in Fig.~\ref{DOS}(f). We can observe two jumps just like in the case of the BI (QVHI) phase. Here, the band gap for the spin-up (spin-down) electrons at the $K$ ($K'$) is significantly large.


\section{Optical conductivities of jacutingaite monolayer} \label{C1}
The spin and valley-polarized optical conductivity components $\sigma_{ij}$ of ML-jacutingaite for both $K$ and $K'$ valleys consist of intra-band (Drude) and inter-band conductivity. The intra-band and inter-band optical conductivity $\sigma_{\alpha\beta}(i \omega, \Delta_{\eta, s})$ can be obtained by Kubo's formula \cite{rodriguez2017casimir,PhysRevB.109.235418}
\begin{equation}\label{a5}
	\sigma_{\alpha\beta}(i \omega, \Delta_{\eta, s})=\sigma_{\alpha\beta}^{intra}(i \omega, \Delta_{\eta, s})+\sigma_{\alpha\beta}^{inter}(i \omega, \Delta_{\eta, s}).
\end{equation}
Here, $(\alpha, \beta = x, y)$. Also note that the frequency $\omega$ is \emph{different} from the frequency of the incident light $\omega_0$ which has been used earlier. The incident light induces a rich band structure whose spectrum is delineated in Equation~\eqref{b1}. Over and above this electromagnetic radiation, an additional in-plane ac electric field of frequency $\omega\ne\omega_0$ may be applied resulting in parallel and transverse (Hall) currents. The response is captured by the optical conductivities. This scheme is, for example, represented in the article~\cite{oka2009photovoltaic}. Furthermore, the presence of dissipation $\Gamma$ allows one to define the complex frequency $\Omega=\omega+i\Gamma$. This complex frequency is used in calculating the conductivities.

At temperature $T=0$ K, the longitudinal and transverse Hall conductivity components are given by \cite{shah2021probing,PhysRevB.109.235418}
\begin{eqnarray}\label{a6}
	\sigma_{x x}^{\text {intra }}\left(i \omega, \Delta_{\eta, s}\right)&=&\frac{\sigma_0}{2 \pi}\frac{4 \mu_{F}^2-\left|\Delta_{\eta, s}\right|^2}{4 \hbar \mu_{F} \Omega} \\ &\times &\Theta\left(2 \mu_{F}-\left|\Delta_{\eta, s}\right|\right), \\
	\sigma_{x x}^{\text {inter }}\left(i \omega, \Delta_{\eta, s}\right)&=&\frac{\sigma_0}{2 \pi}\left(1-\frac{\left|\Delta_{\eta, s}\right|^2}{\hbar^2 \Omega^2}\right) \tan ^{-1}\left(\frac{\hbar \Omega}{\mathcal{M}}\right)\nonumber\\
	&+&\frac{\left|\Delta_{\eta, s}\right|^2}{\hbar \Omega \mathcal{M}},\\
	\sigma_{x y}^{i \text { intra }}\left(i \omega, \Delta_{\eta, s}\right)&=&0,\\
	\sigma_{x y}^{\text {inter }}\left(i \omega, \Delta_{\eta, s}\right)&=&\frac{\sigma_0}{2 \pi}\frac{2 \eta \Delta_{\eta, s}}{\hbar \Omega} \tan ^{-1}\left(\frac{\hbar \Omega}{\mathcal{M}}\right),
\end{eqnarray}
where $\Theta(2\mu_{F}-|\Delta_{\eta, s}|)$ is the Heaviside function and $\mu_{F}$ is the chemical potential. Here, $\sigma_{xx}(i \omega, \Delta_{\eta, s})=\sigma_{yy}(i \omega, \Delta_{\eta, s})$, $\sigma_{yx}(i \omega, \Delta_{\eta, s})=-\sigma_{xy}(i \omega, \Delta_{\eta, s})$, $\sigma_{0}=e^2/4\hbar$, and $\mathcal{M}=\max(|\Delta_{\eta, s}|,2|\mu_{F}|)$ \cite{kort2017topological}. 
\begin{figure*}[ht!]	
\includegraphics[width=0.40\linewidth]{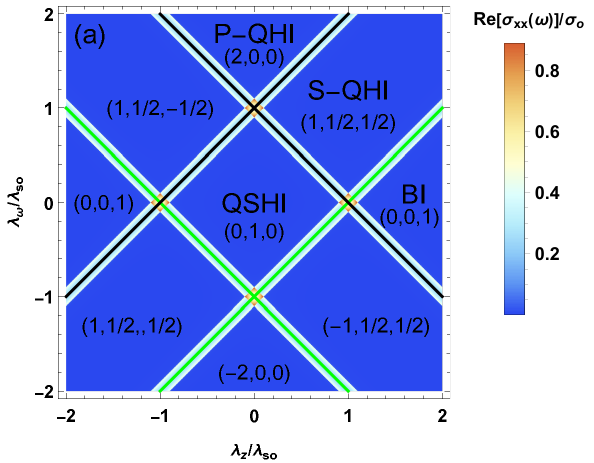}
	\includegraphics[width=0.40\linewidth]{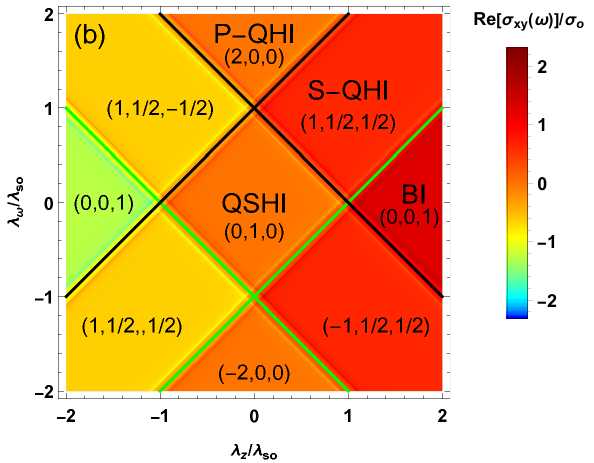}
	\caption{Phase diagram of (a) longitudinal conductivity, (b) transverse Hall conductivity, for ML-jacutingaite in the $\lambda_{z}/\lambda_{so}$ and $\lambda_{\omega}/\lambda_{so}$ plane. The parameters used are, $\hbar \omega=0.2\lambda_{so}$, $\lambda_{so}=81.2$ meV, $\Gamma=0.002\lambda_{so}$, and  $\mu_{F}=0.1\lambda_{so}$.}
	\label{conductivity1}
\end{figure*}

\begin{table*}[t]
	{\caption{Optical transitions in ML-jacutingaite in the $K$ valley in distinct topological regimes. The $\times$ sign represents no transition. The spin-orbit coupling in ML-jacutingaite is $\lambda_{so}$= 81.2 meV. \label{mytab1} }}
	\centering
	\begin{tabular}{lllllll}
		\hline 
		&~~$\lambda_{z}/\lambda_{so}$ &~~~ $\lambda_{\omega}/\lambda_{so}$&~~~~~~~Topological phase&~~~~$T_{1}$&~~~~$T_{2}$      \\ \hline
		&~~~~~~0 &~~~~~~0 &~~~~ QSHI  &~~~~$\hbar\omega/\lambda_{so}(\downarrow)$ &~~~~$\textcolor{red}{\times}(\uparrow)$     \\
		&~~~~~~0 &~~~~~~0.5 &~~~~ QSHI  &~~~~$0.5\hbar\omega/\lambda_{so}(\downarrow)$ &~~~~$1.5\hbar\omega/\lambda_{so}(\uparrow)$      \\
		&~~~~~~0&~~~~~~1 &~~~~ SPM  &~~~~$\textcolor{red}{\times}(\downarrow)$ &~~~~$2\hbar\omega/\lambda_{so}(\uparrow)$          \\
		&~~~~~~1 &~~~~~~0 &~~~~ VSPM  &~~~~$\textcolor{red}{\times}(\downarrow)$ &~~~~$2\hbar\omega/\lambda_{so}(\uparrow)$  \\
		&~~~~~~1.5 &~~~~~~0 &~~~~ BI  &~~~~$0.5\hbar\omega/\lambda_{so}(\downarrow)$ &~~~~$2.5\hbar\omega/\lambda_{so}(\uparrow)$
		
		\\
		&~~~~~~1 &~~~~~~1 &~~~~ P-QHI  &~~~~$0.5\hbar\omega/\lambda_{so}(\downarrow)$ &~~~~$2.5\hbar\omega/\lambda_{so}(\uparrow)$    \\
		&~~~~~~1 &~~~~~~1.5 &~~~~ S-QHI  &~~~~$1.5\hbar\omega/\lambda_{so}(\downarrow)$ &~~~~$3.5\hbar\omega/\lambda_{so}(\uparrow)$    \\ \hline
	\end{tabular}
\end{table*}
Figures~\ref{conductivity1}(a) and (b) display the spin and valley-resolved longitudinal and transverse Hall conductivities in the various topological phases of ML-jacutingaite. The conductivities are plotted as a function of normalized staggered electric potential $\lambda_{z}/\lambda_{so}$ and off-resonant optical field $\lambda_{\omega}/\lambda_{so}$. The
spin and valley-resolved and spin (valley) Chern numbers for different topological phases are also indicated. From Figs.~\ref{conductivity1}(a) and (b), it is obvious that the Dirac mass $\Delta_{\eta, s}$ plays a crucial role in the topological phase transition. As mentioned earlier the phase boundaries appear when $\Delta_{\eta, s}$=0.  The solid black and green lines indicate phase boundaries. The solid black line represents the spin-up and the solid green line represents the spin-down. It is shown that if $\lambda_{z}$
is gradually increased from 0 and $\lambda_{\omega}=0$ remains unchanged, ML-jacutingaite undergoes a topological phase transition from a QSHI to a VSPM and eventually to a BI (QVHI). Similarly, if we fix $\lambda_{z}$=0 and tune the optical field $\lambda_{\omega}$, then the system makes the transition from QSHI  to SPM and finally to P-QHI. If all the processes are adiabatically performed, then, the SPM, VSPM, and SDC phases emerge at the intersection of different topological phases, and these gaps are closed. 
\begin{figure*}[ht!]	
	\includegraphics[width=0.40\linewidth]{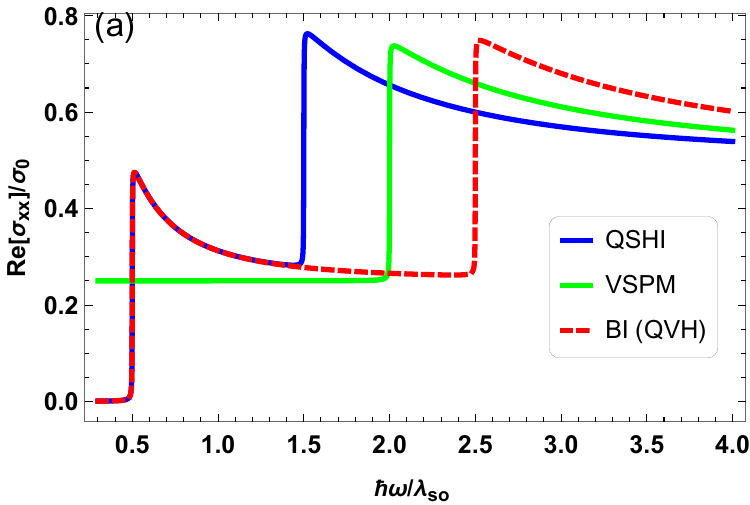}
	\includegraphics[width=0.40\linewidth]{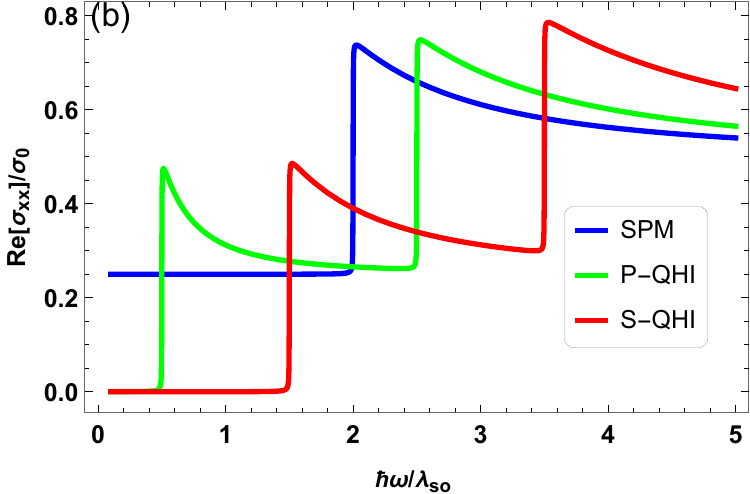}\\
	\includegraphics[width=0.40\linewidth]{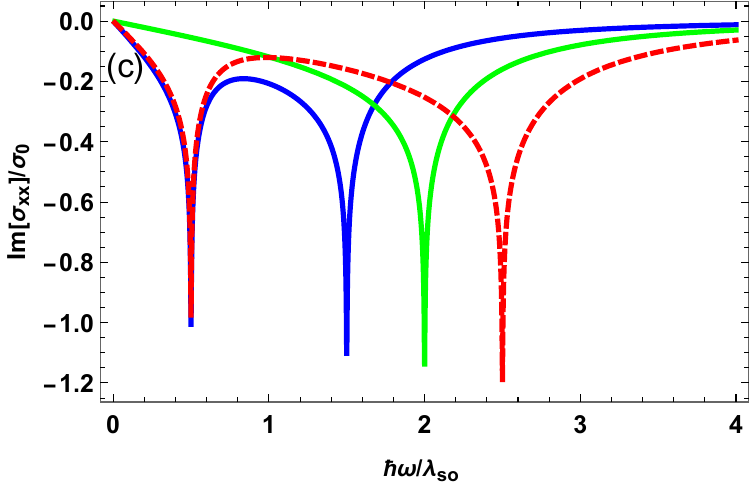}
	\includegraphics[width=0.40\linewidth]{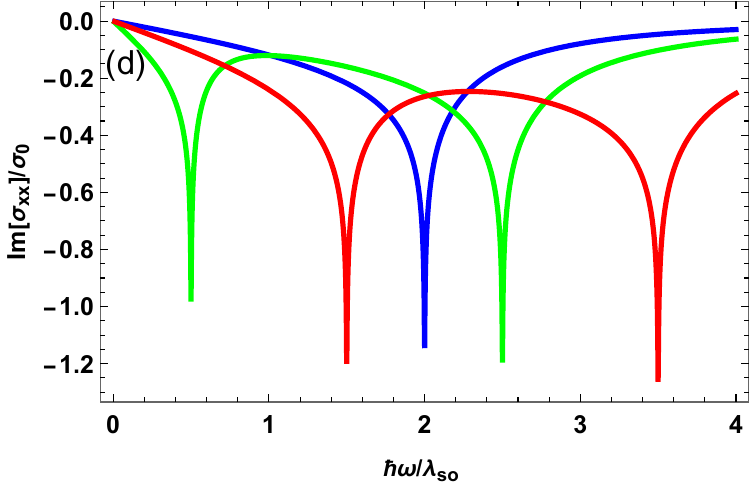}
	\caption{Longitudinal conductivity of jacutingaite as a function of normalized incident photon energy in distinct topological phases. (a) $\mathrm{Re}[\sigma_{xx}]/\sigma_{0}$ in QSHI, VSPM, and BI (QVHI) phases in the $K$ valley. (b) $\mathrm{Re}[\sigma_{xx}]/\sigma_{0}$ in SPM, P-QHI and S-QHI phases in the $K$ valley. (c) $\mathrm{Im}[\sigma_{xx}]/\sigma_{0}$ in QSHI, VSPM, and BI (QVHI) phases in the $K$ valley. (d) $\mathrm{Im}[\sigma_{xx}]/\sigma_{0}$ in SPM, P-QHI and S-QHI phases in the $K$ valley. The parameters used are, $\lambda_{so}=81.2$ meV, $\Gamma=\lambda_{so}\times10^{-3}$, and  $\mu_{F}=0$.}
	\label{conductivity2}
\end{figure*}

\begin{figure*}[ht!]	
\includegraphics[width=0.40\linewidth]{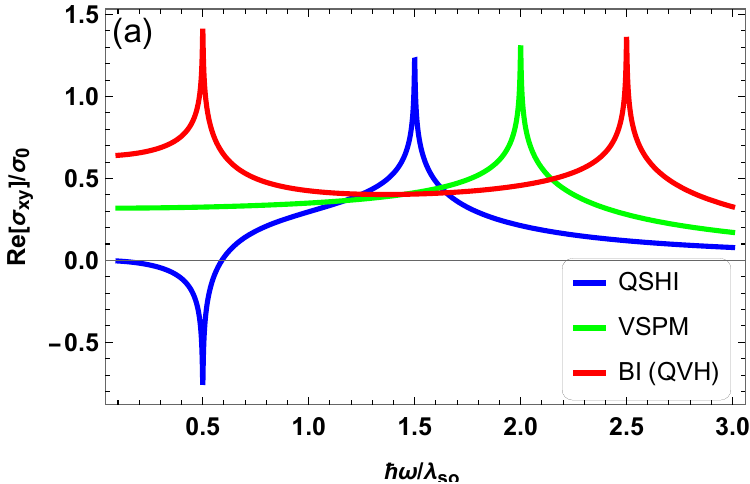}
\includegraphics[width=0.40\linewidth]{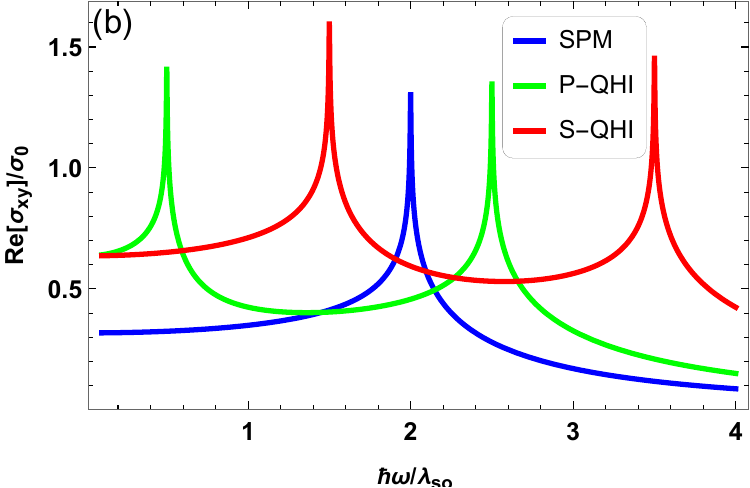}\\	\includegraphics[width=0.40\linewidth]{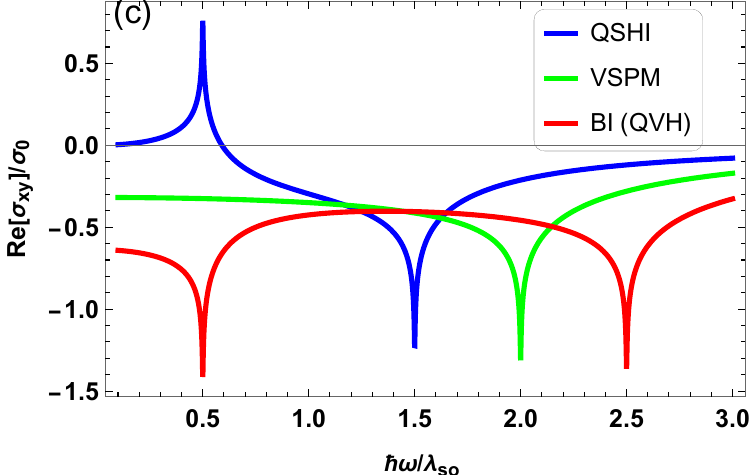}
\includegraphics[width=0.40\linewidth]{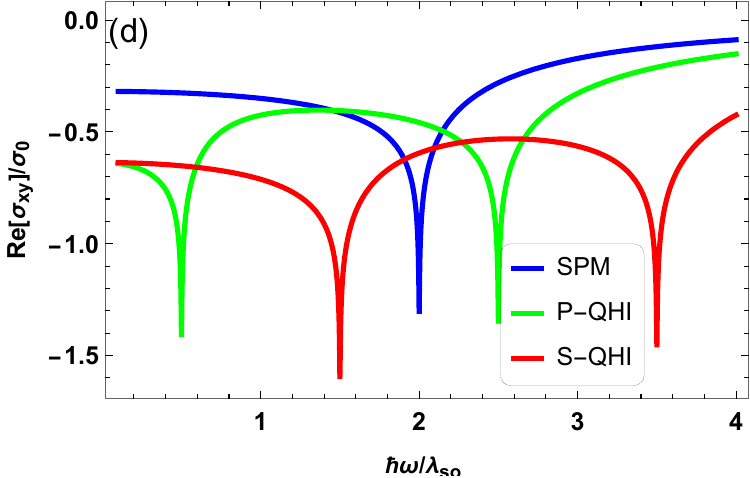}\\
\includegraphics[width=0.40\linewidth]{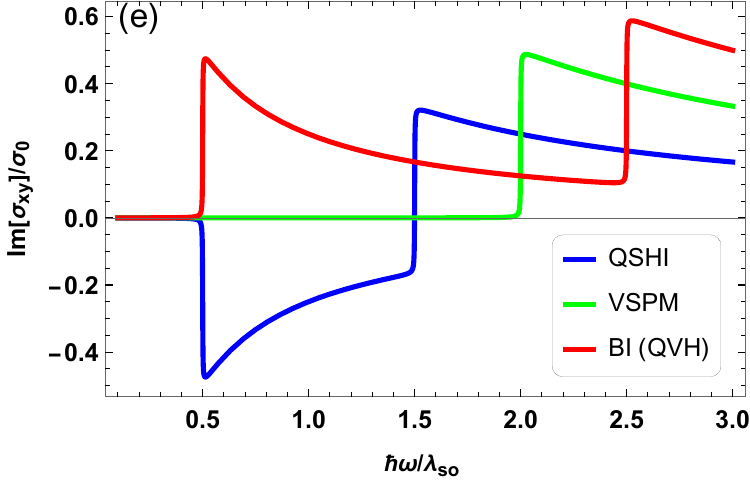}
\includegraphics[width=0.40\linewidth]{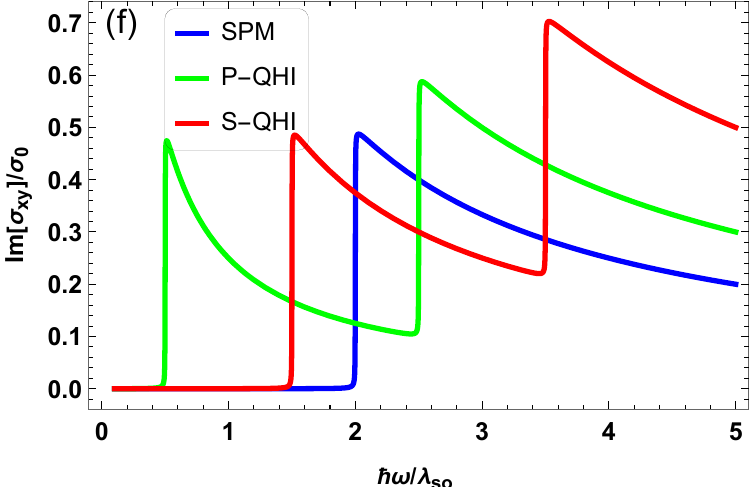}
	\caption{Transverse Hall conductivity of jacutingaite as a function of normalized incident photon energy in distinct topological phases. (a) $\mathrm{Re}[\sigma_{xy}]/\sigma_{0}$ in QSHI, VSPM, and BI (QVHI) phases in the $K$ valley. (b) $\mathrm{Re}[\sigma_{xy}]/\sigma_{0}$ in SPM, P-QHI and S-QHI phases in the $K$ valley. (c) $\mathrm{Re}[\sigma_{xy}]/\sigma_{0}$ in QSHI, VSPM, and BI (QVHI) phases in the $K'$ valley. (d) $\mathrm{Re}[\sigma_{xy}]/\sigma_{0}$ in SPM, P-QHI and S-QHI phases  in the $K'$ valley. (c) $\mathrm{Im}[\sigma_{xy}]/\sigma_{0}$ in QSHI, VSPM, and BI (QVHI) phases in the $K$ valley. (d) $\mathrm{Im}[\sigma_{xy}]/\sigma_{0}$ in SPM, P-QHI and S-QHI phases  in the $K$ valley.  The parameters used are, $\lambda_{so}=81.2$ meV, $\Gamma=\lambda_{so}\times10^{-3}$, and  $\mu_{F}=0$.}
	\label{conductivity3}
\end{figure*}
We now present the results for the spin and valley-resolved longitudinal and transverse Hall conductivities which we shall use to identify signatures of the QSHI, VSPM, BI, SPM, P-QHI, and S-QHI phases. First, we consider the charge-neutral ML-jacutingaite (i.e. $\mu_{F}=$0). In Figs. \ref{conductivity1}(a)-(d) we show the spin-valley resolved longitudinal conductivities as a function of normalized photonic energies at zero temperatures
($T$ = 0) for various values of electric and optical fields in the $K$ valley. For these simulations, we chose
chemical potential to be $\mu_{F}=\lambda_{so}$ with $\lambda_{so}=81.2$ meV. It must be noted that the interband and intraband transitions can only occur between bands of the
same spin index. In Fig.~\ref{conductivity1}(a), we demonstrate the conductivity spectra for three topological phases namely, QSHI, VSPM, and BI respectively.  In the QSHI phase ($\lambda_{z}=0.5\lambda_{so}$, $\lambda_{\omega}=0$), the degeneracy between gaps for the spin-up and spin-down electrons is broken (refer toFig.~\ref{Bandgaps}(a) again). Two jumps/peaks originated from the spin-up and spin-down interband optical transitions can be seen in the conductivity spectra (shown by the blue curve). The optical excitation energies corresponding to the spin-up and spin-down optical transitions are $0.5\lambda_{so}$ and $1.5\lambda_{so}$ respectively. We have displayed the optical response of the ML-jacutingaite in the VSPM ($\lambda_{z}=\lambda_{so}$, $\lambda_{\omega}=0$) state by red curve in Fig.~\ref{conductivity1}(a). According to Fig.~\ref{Bandgaps}(b), the gap of the spin-down electron bands is closing and results in a Dirac point. In this phase, there is a single jump in the conductivity at $\hbar\omega=2\lambda_{so}$ associated with spin-up electron transitions between the gapped bands. The staggered electric field pushes $\sigma_{xx}$ to higher photonic energies without making any changes in its sign.  By increasing the strength of the staggered potential the ML-jacutingaite system transitions from the QSHI to the BI (QVHI) phase. The optical response ascribed to the spin-up and spin-down electrons in the BI regime is shown by the dashed red curve in Fig.~\ref{conductivity1}(a). In the BI (QVHI) phase, the lowest band gap is opened again and the feature splits into two spin-polarized jumps as shown in Fig.~\ref{conductivity1}(a). The optical excitation energies of the spin-up and spin-down electrons in this phase are determined as $\hbar\omega=0.5\lambda_{so}$ and $\hbar\omega=2.5\lambda_{so}$ respectively.

Next, we display the spin-valley dependent longitudinal conductivities as a function of normalized photonic energies for the $K$ valley in the SPM, P-QHI, and S-QHI regimes in Fig.~\ref{conductivity1}(b). When $\lambda_{z}=0$, $\lambda_{\omega}=\lambda_{so}$ is applied, the gap closes for spin-down (spin-up) state at $K$ ($K'$) valley, resulting in a SPM state as shown in Fig.~\ref{Bandgaps}(c). The longitudinal conductivity of the SPM state is shown by a blue curve as presented in Fig.~\ref{conductivity1}(b). Again, we can see one resonant jump due to the spin-up optical transition at $\hbar\omega=2\lambda_{so}$. By varying the off-resonant optical field strength, the system reaches the P-QHI $\lambda_{z}=0$, $\lambda_{\omega}=1.5\lambda_{so}$ phase. The green curve shows the optical response of the spin-up and spin-down at the $K$ valley in the P-QHI phase. The spin-up and spin-down resonant jumps can be seen at the optical transition energies $0.5\lambda_{so}$ and $\hbar\omega=2.5\lambda_{so}$ respectively. In Fig.~\ref{conductivity3}(c) and (d), we plot the imaginary parts of longitudinal conductivity for each spin in the $K$ valley for the QSHI, VSPM, BI, SPM, P-QHI, and S-QHI phases, where the signs of $\sigma_{xx}(i \omega, \Delta_{\eta, s})$ are negative. 

\begin{figure*}[t!]
	\centering
	\includegraphics[width=0.40\linewidth]{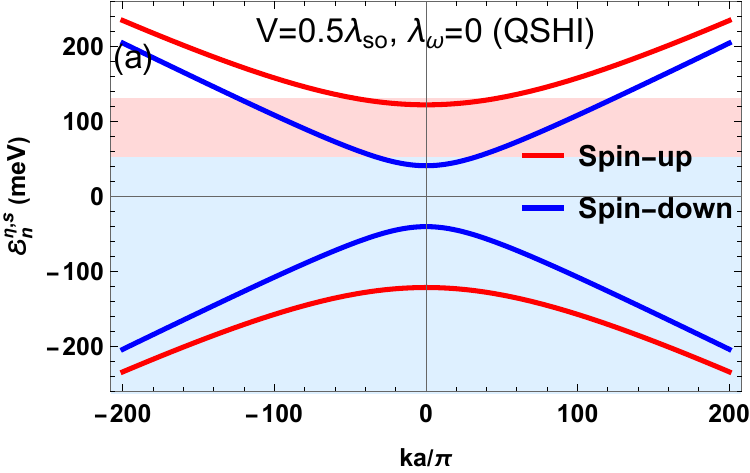}
	\includegraphics[width=0.40\linewidth]{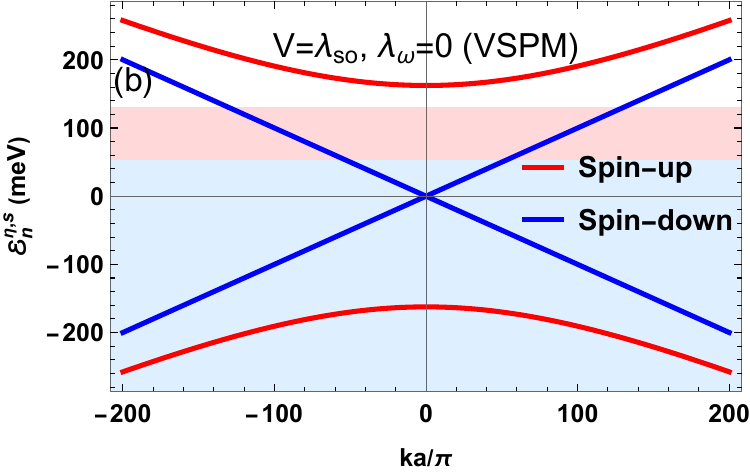}\\
	\includegraphics[width=0.40\linewidth]{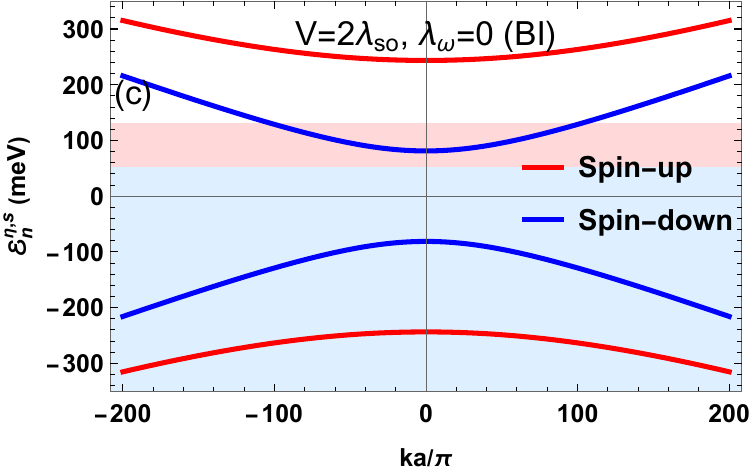}
	\includegraphics[width=0.40\linewidth]{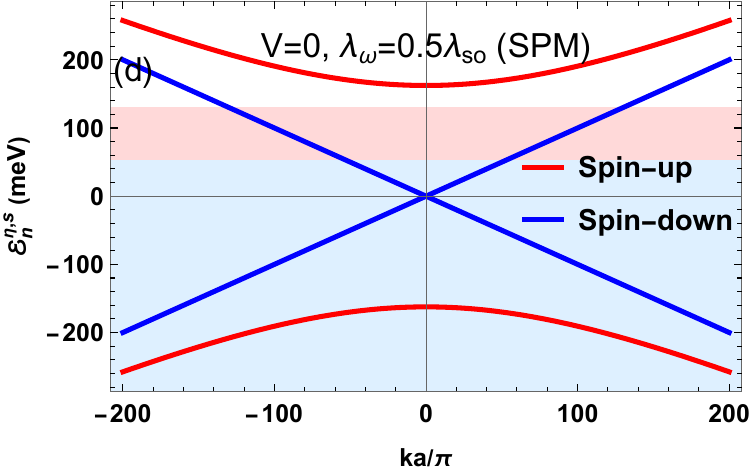}\\
	\includegraphics[width=0.40\linewidth]{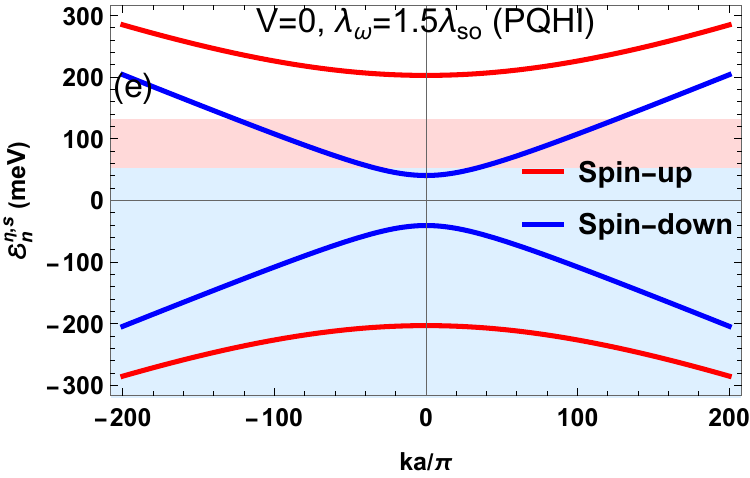}
	\includegraphics[width=0.40\linewidth]{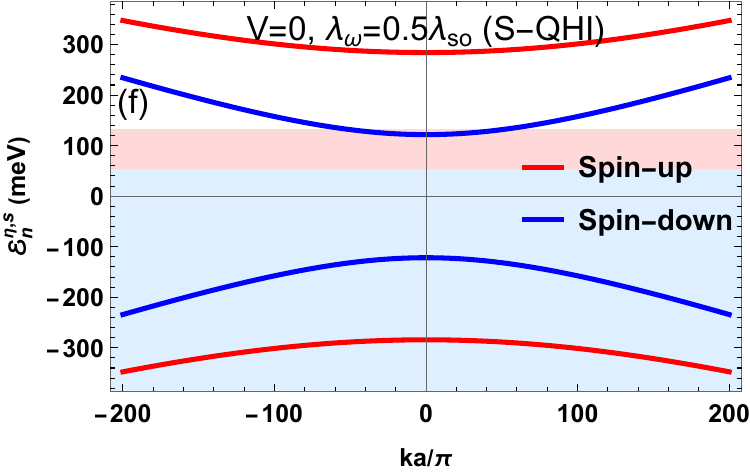}
	\caption{Band structure of jacutingaite in distinct topological phases at the $K$ valley. (a) QSHI ($\lambda_{z}=0.5\lambda_{so}$, $\lambda_{\omega}=0$), (b) VSPM  ($\lambda_{z}=\lambda_{so}$, $\lambda_{\omega}=0$), (c) BI ($\lambda_{z}=1.5\lambda_{so}$,  $\lambda_{\omega}=0$), (d) SPM ($\lambda_{z}=0$,  $\lambda_{\omega}=\lambda_{so}$), (e) P-QHI ($\lambda_{z}=0$, $\lambda_{\omega}=1.5\lambda_{so}$) and (f) S-QHI ($\lambda_{z}=\lambda_{so}$,  $\lambda_{\omega}=1.5\lambda_{so}$) respectively. The red and blue curves refer to spin-up and spin-down energy bands, respectively. The blue and red-shaded region indicates the level of chemical potentials $\mu_{F}$.}
	\label{Bandgaps1}
\end{figure*}

In the following, we aim to investigate the
spin and valley Hall effects of ML-jacutingaite. The spin Hall effect
(SHE) is an analog to the charge based electronic Hall effect. In SHE spin-up and spin-down electrons flow to the opposite edges of semiconductors when a longitudinal electric field is applied \cite{wunderlich2004experimental,kato2004observation}. The active control and manipulation of spin degrees of freedom is at the core of the spintronics pprogram. ML-jacutingaite is predicted to become one of the very first large-gap Kane-Mele quantum spin Hall insulators and and can be vouched as a promising material for spintronic applications \cite{marrazzo2018prediction}. Similarly, in the valley Hall effect electrons from different valleys experience opposite Lorentz-like forces and flow to opposite transverse edges. This allows for the possibility of potential valleytronic devices that can be utilized to encode information \cite{schaibley2016valleytronics}. We now analyze the finite frequency spin and valley Hall conductivities in distinct topological phases.

In Fig. \ref{conductivity3}(a),  the spin and valley Hall conductivity as a function of the dimensionless photonic energy for the charge-neutral ($\mu_{F}$ = 0) system in distinct topological phases is presented. In the QSHI phase, we can see the two sharp peaks originating from spin-up and spin-down electron optical transitions in the conductivity spectra. It should be noted that a positive conductivity peak indicates a net spin-up (-down) accumulation in one transverse direction while a negative conductivity feature produces a net spin-up (-down) accumulation on the opposite edge. The first negative (positive) peak occurs at $\hbar\omega=0.5\lambda_{so}$ and is associated with the excitation of the spin-down (spin-up) electron while the second positive (negative) feature can be seen at $\hbar\omega=1.5\lambda_{so}$ due to the excitation of the spin-up (spin-down) electron in the $K$ ($K'$) valley. These are purely the interband transitions. In the VSPM state, the spin and valley Hall conductivities are shown in Figs. \ref{conductivity3}(a) and (c). Only one feature can be seen originating from the excitation of the spin-up (spin-down) electron at $\hbar\omega=2\lambda_{so}$ the $K$ ($K'$) valley. In this phase, the electrons of a specific valley label always flow to one transverse edge. As we further increase the strength of the staggered sublattice potential the system transitions into the BI (QVHI) phase, where the second gap reopening leads to the reappearance of two peaks in the conductivity spectra. Conversely, in the BI (QVHI) phase, electrons from a particular valley move in only one direction for all photonic energies. Again, we can observe two sharp features associated with spin-up and spin-down electron interband transitions for the $K$ and $K'$ valleys as presented in Figs. \ref{conductivity3}(a) and (c) respectively. In this phase, the peak originated by spin-down (spin-up) optical transition moves to higher (lower) energy, and the second spin-up (spin-down) peak to lower (higher) energy in the $K$ ($K'$) valley.

In the SPM phase, a single peak reappears as in the VSPM case [see Figs. \ref{conductivity3}(b) and (d)]. The spin and valley Hall conductivities change their signs in the respective valley.
In the PS-QHI and S-QHI phases, spin and valley Hall conductivities peaks are always positive (negative) for spin-up and spin-down electron optical transitions as illustrated in Figs. \ref{conductivity3}(b) and (d) for the $K$ ($K'$) valley. We can observe two peaks at different photon energies. As the band gap of the ML-jacutingaite is increasing the resonant peaks are moving towards the right in the photonic energies. The optical excitation energies corresponding to different spin-up and spin-down optical transitions are tabulated in Table \ref{mytab1} for distinct topological phases. In the same way, we have plotted the imaginary part of the spin and valley Hall conductivities in the aforementioned phases in Figs. \ref{conductivity3}(e) and (f) at the $K$ valley only. Figure \ref{conductivity3}(e) shows calculations for the $K$ valley in the QSHI, VSPM, and BI (QVHI) phases for spin-up and spin-down optical excitations. In the QSHI phase, negative and positive jumps can be seen at the excitation photonic energies. The negative and positive jumps are originated by the spin-down and spin-up optical transitions in the $K$ valley respectively. Similarly, for the VSPM and BI (QVHI) phases, there are one and two resonant jumps respectively.  In the SPM, P-QHI, and S-QHI phases the sign of the imaginary Hall conductivity does not change by increasing the electric field and irradiated optical field. The external stimuli shift the resonant jumps to higher energy as depicted in Fig. \ref{conductivity3}(f).

\begin{figure*}[ht!]	
	\includegraphics[width=0.40\linewidth]{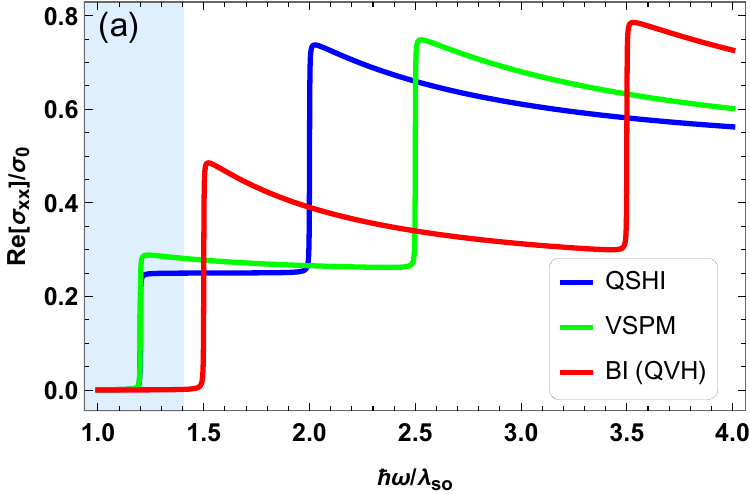}
	\includegraphics[width=0.40\linewidth]{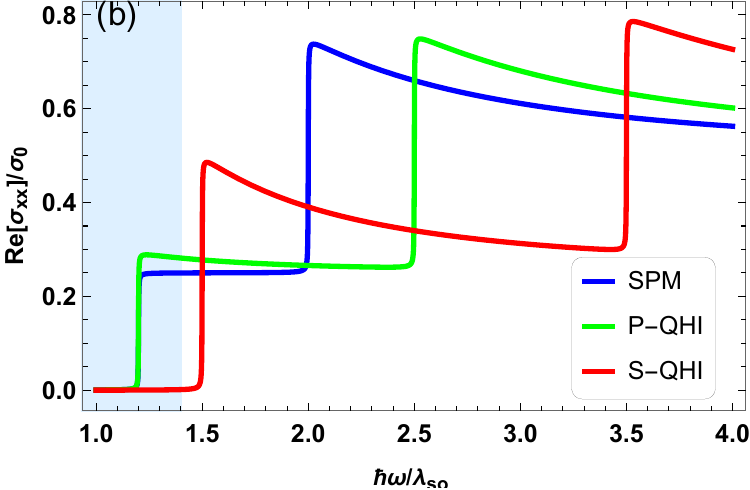}\\
	\includegraphics[width=0.40\linewidth]{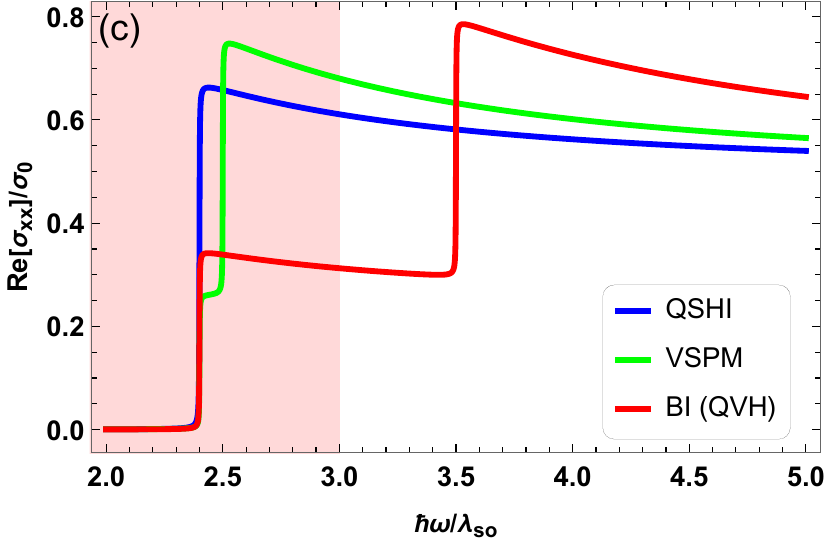}
	\includegraphics[width=0.40\linewidth]{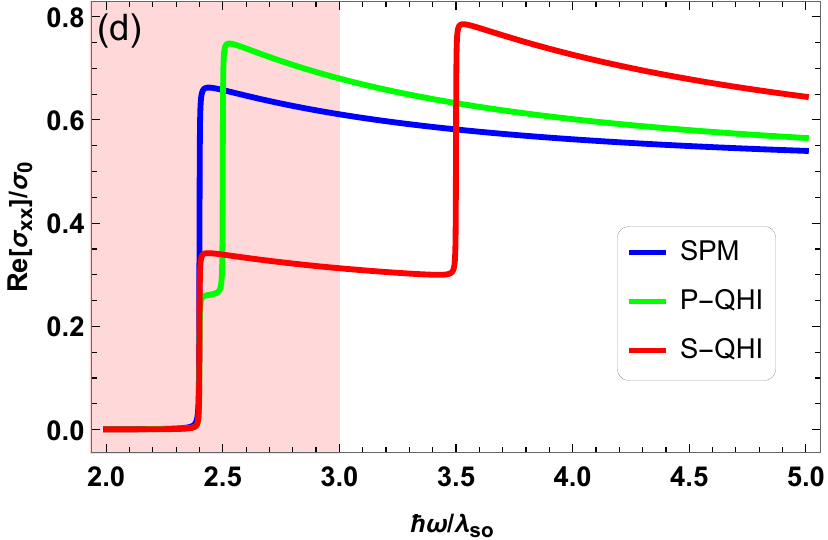}
	\caption{Longitudinal conductivity of jacutingaite as a function of normalized incident photon energy in distinct topological phases. (a) $\mathrm{Re}[\sigma_{xx}]/\sigma_{0}$ in the (a) QSHI, VSPM, and BI (QVHI)  (b) SPM, P-QHI, and S-QHI phases in the $K$ valley for a fixed chemical potential $\mu_{F}=0.6\lambda_{so}$. (c) $\mathrm{Re}[\sigma_{xx}]/\sigma_{0}$ in the (a) QSHI, VSPM, and BI (QVHI)  (d) SPM, P-QHI and S-QHI phases in the $K$ valley for a fixed chemical potential $\mu_{F}=1.6\lambda_{so}$. The parameters used are, $\lambda_{so}=81.2$ meV and $\Gamma=\lambda_{so}\times10^{-3}$. The blue and red-shaded region indicates the level of chemical potentials $\mu_{F}$.}
	\label{conductivity5}
\end{figure*}

It is instructive to discuss the effect of controlling the longitudinal conductivity spectra in distinct topological regimes by varying the chemical potential of ML-jacutingaite. For finite charge doping, we have shown the real part of the spin and valley-resolved longitudinal conductivities in Figs. \ref{conductivity5}(a)-(d). For illustration purposes, we consider two different values of chemical potentials
$\mu_{F}=0.6\lambda_{so}$ and $\mu_{F}=1.6\lambda_{so}$ in in distinct topological phases. To demonstrate how various chemical potentials impact the longitudinal conductivities, we present a detailed analysis of band structures of ML-jacutingaite material in the $K$ valley in distinct topological phases in Figs.~\ref{Bandgaps1}(a)-(f). The blue and red shaded regions indicate the limit of $\mu_{F}=0.6\lambda_{so}$ and $\mu_{F}=1.6\lambda_{so}$ respectively. In the first case, the chemical potential is 
$\mu_{F}=0.6\lambda_{so}$. It should be kept in mind that the effect of doping can be neglected in regions where $|\mu_{F}|<\left|\Delta_{\eta, s}\right|$. The longitudinal conductivities are similar to those of neutral ML-jacutingaite. However, close to phase transition boundaries where $|\mu_{F}|>\left|\Delta_{\eta, s}\right|$, intraband optical transitions occur and the influence of doping becomes significant. In contrast to neutral jacutingaite layers, for example, a jump in $\sigma_{xx}$ emerges around $\hbar\omega=0.5 \lambda_{so}$ for $\mu_{F}=0.001\lambda_{so}$. Further increasing the chemical potential allows us to shift the jump's position to lower values of photon energies in certain topological phases. In the QSHI phase, $\mu_{F}$ residing in between the spin-up and spin-down conduction bands as depicted in Fig.~\ref{Bandgaps1}(a). In the longitudinal optical conductivity spectra, we can observe two jumps in the QSHI phase [see Fig.~\ref{conductivity5}(a)]. If $\hbar\omega<2 \mu_{F}$,  then the spin-down interband transitions are forbidden due to the Pauli exclusion principle. The interband transitions of the spin-down electrons now become Pauli-blocked. The first jump originated by the spin-down electron is due to the intraband optical transition, while the second jump is associated with purely interband transition due to the spin-up electron. The intraband transitions are shown in the blue shaded region in Fig.~\ref{conductivity5}(a).   In contrast to neutral jacutingaite (where only one jump existed for spin-up optical transitions see Fig.~\ref{conductivity1}(a)), in the VSPM phase, two jumps can be seen as shown in Fig.~\ref{conductivity5}(a).  In this scenario, only the intraband transitions of the spin-down electrons occur. According to the optical selection rules, the spin-up interband transitions are allowed in this regime as shown in Fig.~\ref{conductivity5}(a). In contrast to neutral jacutingaite in the VSPM phase, where only one jump existed for spin-up optical transitions as shown in Fig.~\ref{conductivity1}(a).

In the BI (QVHI) phase, the band gap for spin-up electrons is significantly large as shown in Fig.~\ref{Bandgaps1}(c). Due to the substantial energy separation between the valence and conduction bands, we have only interband transitions for spin-up and spin-down electrons and intraband transitions no longer occur. For $\mu_{F}=0.6\lambda_{so}$,  similar behavior is seen in the case of undoped jacutingaite in the BI (QVHI) phase (shown Fig.~\ref{conductivity5}(a)).  Figure.~\ref{conductivity5}(b) depicts the impact of altering the chemical potential on the longitudinal conductivity spectra in the SPM, P-QHI, and S-QHI topological regimes. For the SPM and P-QHI phases, both intraband and interband transitions occur. In the S-QHI phase, the band gap is very large for spin-up and spin-down electrons as presented in Fig.~\ref{Bandgaps1}(f). The jumps in the spin-valley resolved longitudinal conductivities are due to spin-up and spin-down optical interband transitions as displayed in Fig.~\ref{conductivity5}(b). In Figs.~\ref{conductivity5}(c) and (d), we plotted the real part of the longitudinal conductivity as a function of photon energies for $\mu_{F}=1.6\lambda_{so}$ in distinct topological phases.  
The jumps originated by the intraband optical transitions are shown in the red-shaded region. One can observe that the interband optical transitions due to the spin-down electrons are completely Pauli-blocked. All the optical transitions due to the spin-down electrons are intraband transitions in the QSHI, VSPM, and BI (QVHI) phases as shown in Fig.~\ref{conductivity5}(c). Due to the larger bandgap in the BI (QVHI) phase, we have only one interband optical transition for spin-up electrons. Similar behaviors can be observed in the SMP, P-QHI, and S-QHI phases as depicted in Fig.~\ref{conductivity5}(d). 

By utilizing the optical conductivity of monolayer jacutingaite, one can calculate the absorption spectra by solving the Maxwell equations with a boundary conditions. The imaginary part of the Hall conductivity can be used to determine the circular dichroism and valley polarization in these 2D materials. Moreover, the conductivity spectra can benefit to an in-depth understanding of the  Faraday and Kerr responses of monolayer jacutingaite which can be potentially applied for magneto-optic, spintronic, and valleytronic devices in the THz range.  The optical conductivity of monolayer jacutingaite is a valuable tool for identifying distinct topological phases in these 2D materials. The band gaps of the spin-up and spin-down electrons in different topological phases can be experimentally measured
with reasonable accuracy. 

\section{Conclusions} \label{E1}
In conclusion, we started with the effective Hamiltonian of monolayer jacutingaite subjected to a circularly polarized optical field and staggered sublattice potential to derive its band structure. After diagonalization, we investigated the exceptionally rich spin and valley-resolved phase diagram of the system, through calculation of Berry curvatures and Chern numbers. We found that the system can transition from the QSHI state to the trivial BI (QVHI) state by applying a staggered sublattice potential. On the other hand, for a fixed electric field, as the circularly polarized optical field gradually increases from zero, the system transitions from the QSHI phase to the SPM and further to the P-QHI phase. This rich phase structure can be validated by circular dichroism studies. 

Furthermore, we computed the spin and valley polarized conductivities in all the posited topological phases. Our analysis shows that the spin- and valley-Hall conductivities exhibit strong topological  dependence. Finally, the impact of chemical potentials is investigated. Most of the interband transitions of the spin-down electrons are blocked due to the Pauli exclusion principle in the condition of high chemical potential. Overall the study provides a reliable guideline for designing magneto-optic, spintronic, and valleytronic devices in the THz range.

\section{acknowledgments}
M. Shah acknowledges  financial support from the postdoctoral research grant YS304023905. 
\renewcommand{\bibname}{References}

\bibliography{References}
\end{document}